\documentclass{emulateapj}
\usepackage{graphicx}
\usepackage{txfonts}
\usepackage{natbib}
\usepackage[scaled]{helvet}
\usepackage{epsfig}
\usepackage{url}
\usepackage[normalem]{ulem}
\usepackage{color}
\bibpunct{(}{)}{;}{a}{}{,}
\newcommand\mynotes[1]{\textcolor{red}{#1}} 
\newcommand{\MP}{\textit{M}$_{\textup{p}}$} 
\newcommand{\RP}{\textit{R}$_{\textup{p}}$} 
\newcommand{\MJ}{\textit{M}$_{\textup{J}}$} 
\newcommand{\RJ}{\textit{R}$_{\textup{J}}$} 
\newcommand{\bjdtdb}{\ensuremath{\rm {BJD_{TDB}}}}
\newcommand{\feh}{\ensuremath{\left[{\rm Fe}/{\rm H}\right]}}
\newcommand{\teff}{\ensuremath{T_{\rm eff}}}
\newcommand{\ecosw}{\ensuremath{e\cos{\omega_*}}}
\newcommand{\esinw}{\ensuremath{e\sin{\omega_*}}}
\newcommand{\msun}{\ensuremath{\,M_\Sun}}
\newcommand{\rsun}{\ensuremath{\,R_\Sun}}
\newcommand{\lsun}{\ensuremath{\,L_\Sun}}
\newcommand{\mj}{\ensuremath{\,M_{\rm J}}}
\newcommand{\rj}{\ensuremath{\,R_{\rm J}}}
\newcommand{\fave}{\langle F \rangle}
\newcommand{\fluxcgs}{10$^9$ erg s$^{-1}$ cm$^{-2}$}
\newcommand{\ms}{\,m\,s$^{-1}$} 
\newcommand{\kms}{\,km\,s$^{-1}$} 

\begin{document}

\title{\bf KELT-14\lowercase{b} and KELT-15\lowercase{b}: An Independent Discovery of WASP-122\lowercase{b} and a New Hot Jupiter }

\author{
Joseph E. Rodriguez$^1$, 
Knicole D. Col\'on$^{2,3}$,
Keivan G. Stassun$^{1,4}$, 
Duncan Wright$^{5,6}$,
Phillip A. Cargile$^{7}$, 
Daniel Bayliss$^{8}$, 
Joshua Pepper$^{9}$, 
Karen A. Collins$^{1}$,
Rudolf B. Kuhn$^{10}$, 
Michael B. Lund$^1$, 
Robert J. Siverd$^{11}$, 
George Zhou$^{7,12}$, 
B. Scott Gaudi$^{13}$,
C.G. Tinney$^{5,6}$,
Kaloyan Penev$^{14}$,
T.G. Tan$^{15}$, 
Chris Stockdale$^{16,17}$,
Ivan A. Curtis$^{18}$, 
David James$^{19}$, 
Stephane Udry$^{8}$,
Damien Segransan$^{8}$,
Allyson Bieryla$^{7}$,
David W. Latham$^{7}$,
Thomas G. Beatty$^{20,21}$,
Jason D. Eastman$^{7}$,
Gordon Myers$^{16,22}$,
Jonathan Bartz$^{9}$,
Joao Bento$^{12}$,
Eric L. N. Jensen$^{23}$,
Thomas E. Oberst$^{24}$,
Daniel J. Stevens$^{13}$
}

\affil{$^1$Department of Physics and Astronomy, Vanderbilt University, 6301 Stevenson Center, Nashville, TN 37235, USA}
\affil{$^2$NASA Ames Research Center, M/S 244-30, Moffett Field, CA 94035, USA}
\affil{$^3$Bay Area Environmental Research Institute, 625 2nd St. Ste 209 Petaluma, CA 94952, USA}
\affil{$^4$Department of Physics, Fisk University, 1000 17th Avenue North, Nashville, TN 37208, USA}
\affil{$^{5}$School of Physics, UNSW Australia, 2052. Australia}
\affil{$^{6}$Australian Centre for Astrobiology, UNSW Australia, 2052. Australia}
\affil{$^7$Harvard-Smithsonian Center for Astrophysics, 60 Garden St, Cambridge, MA 02138, USA}
\affil{$^{8}$Observatoire Astronomique de l'Universit\'e de Gen\`eve, Chemin des Maillettes 51, 1290 Sauverny, Switzerland}
\affil{$^9$Department of Physics, Lehigh University, 16 Memorial Drive East, Bethlehem, PA 18015, USA}
\affil{$^{10}$South African Astronomical Observatory, PO Box 9, Observatory 7935, South Africa}
\affil{$^{11}$Las Cumbres Observatory Global Telescope Network, 6740 Cortona Dr., Suite 102, Santa Barbara, CA 93117, USA}
\affil{$^{12}$Research School of Astronomy and Astrophysics, Australian National University, Canberra, ACT 2611, Australia}
\affil{$^{13}$Department of Astronomy, The Ohio State University, Columbus, OH 43210, USA}
\affil{$^{14}$Department of Astrophysical Sciences, Princeton University, Princeton, NJ 08544, USA}
\affil{$^{15}$Perth Exoplanet Survey Telescope, Perth, Australia}
\affil{$^{16}$American Association of Variable Star Observers, 49 Bay State Rd., Cambridge, MA 02138, USA}
\affil{$^{17}$Hazelwood Observatory}
\affil{$^{18}$Adelaide, Australia}
\affil{$^{19}$Cerro Tololo InterAmerican Observatory, Casilla 603, La Serena, Chile}
\affil{$^{20}$Department of Astronomy \& Astrophysics, The Pennsylvania State University, 525 Davey Lab, University Park, PA 16802}
\affil{$^{21}$Center for Exoplanets and Habitable Worlds, The Pennsylvania State University, 525 Davey Lab, University Park, PA 16802}
\affil{$^{22}$5 Inverness Way, Hillsborough, CA 94010, USA}
\affil{$^{23}$Department of Physics and Astronomy, Swarthmore College, Swarthmore, PA 19081, USA}
\affil{$^{24}$Department of Physics, Westminster College, New Wilmington, PA 16172, USA}


\shorttitle{KELT-14b and KELT-15b}

\begin{abstract}
We report the discovery of KELT-14b and KELT-15b, two hot Jupiters from the KELT-South survey. KELT-14b, an independent discovery of the recently announced WASP-122b, is an inflated Jupiter mass planet that orbits a $\sim5.0^{+0.3}_{-0.7}$ Gyr, $V$ = 11.0, G2 star that is near the main sequence turnoff. The host star, KELT-14 (TYC 7638-981-1), has an inferred mass $M_{*}$~=~$1.18_{-0.07}^{+0.05}$~$M_{\sun}$ and radius $R_{*}$~=~$1.37\pm{-0.08}$~$R_{\sun}$, and has \teff~=~$5802_{-92}^{+95}$~K, $\log{g_*}$~=~$4.23_{-0.04}^{+0.05}$ and \feh~=~$0.33\pm{-0.09}$. The planet orbits with a period of $1.7100588 \pm 0.0000025$ days ($T_{0}$=2457091.02863$\pm$0.00047) and has a radius \RP~=~$1.52_{-0.11}^{+0.12}$~\RJ\space and mass \MP~=~$1.196\pm0.072$~\MJ, and the eccentricity is consistent with zero. KELT-15b is another inflated Jupiter mass planet that orbits a $\sim$ $4.6^{+0.5}_{-0.4}$ Gyr, $V$ = 11.2, G0 star (TYC 8146-86-1) that is near the ``blue hook" stage of evolution prior to the Hertzsprung gap, and has an inferred mass $M_{*}$~=~$1.181_{-0.050}^{+0.051}$~$M_{\sun}$ and radius $R_{*}$~=~$1.48_{-0.04}^{+0.09}$~$R_{\sun}$, and \teff~=~$6003_{-52}^{+56}$~K, $\log{g_*}$~=~$4.17_{-0.04}^{+0.02}$ and \feh~=~$0.05\pm0.03$. The planet orbits on a period of $3.329441 \pm 0.000016$ days ($T_{0}$ = 2457029.1663$\pm$0.0073) and has a radius \RP~=~$1.443_{-0.057}^{+0.11}$~\RJ\space and mass \MP~=~$0.91_{-0.22}^{+0.21}$~\MJ\space and an eccentricity consistent with zero. KELT-14b has the second largest expected emission signal in the K-band for known transiting planets brighter than $K<10.5$. Both KELT-14b and KELT-15b are predicted to have large enough emission signals that their secondary eclipses should be detectable using ground-based observatories.

\end{abstract}

\section{\bf{Introduction}}

The confirmation of over 1000 transiting exoplanets to date is due to the success of ground-based photometric surveys such as HATNet \citep{Bakos:2004}, SuperWASP \citep{Pollacco:2006}, XO \citep{McCullough:2006}, and TrES \citep{Alonso:2004}, and the space-based missions CoRoT \citep{Baglin:2006} and Kepler \citep{Borucki:2010}.  The field has shifted from pure discovery to understanding the demographics of exoplanets and atmospheric characterization. However, many of the discovered planets are too faint or too small for performing atmospheric characterization with current facilities. To date, there are only 29 giant transiting planets orbiting stars with $V$ $<$ 11.5 in the southern hemisphere\footnote{www.exoplanets.org, as of September 2015}. 

It is believed that ``hot Jupiters," gas giant planets that orbit extremely close (orbital periods of a few days) to their host stars, must form beyond the ``Snow Line.'' Once formed, the giant planets can migrate inward through various methods \citep{Tanaka:2002, Masset:2003, D'Angelo:2008, Jackson:2008, Cloutier:2013}.  It has been proposed that Jupiter experienced migration early in its lifetime, but did not migrate all the way inward due to the gravitational pull of Saturn \citep{Walsh:2011}. These hot Jupiters, specifically ones orbiting solar-like stars, provide insight into alternate evolutionary scenarios. 

The Kilodegree Extremely Little Telescope (KELT) exoplanet survey, operated and owned by Vanderbilt University, Ohio State University, and Lehigh University, has been observing $>$ 60$\%$ of the sky with a cadence of 10 to 20 minutes for many years. The project uses two telescopes, KELT-North at Winer Observatory in Sonoita, Arizona and KELT-South at the South African Astronomical Observatory (SAAO) in Sutherland, South Africa. The survey is optimized for high-precision ($\le$1\% RMS) photometry for stars with 8 $\le$ $V$ $\le$ 11 to enable transit discovery of giant planets. Each telescope has a 42 mm aperture, 26$^{\circ}\times26^{\circ}$ field of view, and a pixel scale of 23$\arcsec$/pixel \citep{Pepper:2007, Pepper:2012}. The first telescope in the survey, KELT-North, has announced six planets orbiting stars brighter than $V$ = 11 \citep{Siverd:2012,Beatty:2012,Pepper:2013,Collins:2014,Bieryla:2015,Fulton:2015}. The younger counterpart in the survey, KELT-South, has already announced one planet, KELT-10b \citep{Kuhn:2015}.

In this paper, we present the discovery of a new hot Jupiter by KELT-South, which we name KELT-15b.  We also present another hot Jupiter, which we refer to in this paper as KELT-14b.  Shortly before the completion of this paper, a draft manuscript was posted to the arXiv \citep{Turner:2015} describing the discovery of three new exoplanets by the SuperWASP survey.  One of the planets they name WASP-122b, which is the same planet we designate as KELT-14b.  Since the data we present in this paper were collected independently and the analysis performed before the announcement of WASP-122b, we have chosen to discuss our findings as an independent discovery of this planet, and we refer to it here as KELT-14b.  However, we acknowledge the prior announcement of it as WASP-122b.

The paper is organized as follows with each section including both discovered systems, KELT-14b and KELT-15b. In \S2 we present our discovery and follow-up observations (photometric and spectroscopic). We present our stellar characterization analysis and results in \S3. The global modeling and resulting planetary parameters are discussed in \S4 with our false positive analysis described in \S5. In \S6 we describe the evolutionary analysis, long-term follow-up to look for additional companions in each system, and the value each planetary system has for future atmospheric characterization observations. We summarize our results and conclusions in \S7.

\begin{figure*}[ht]
\centering\epsfig{file=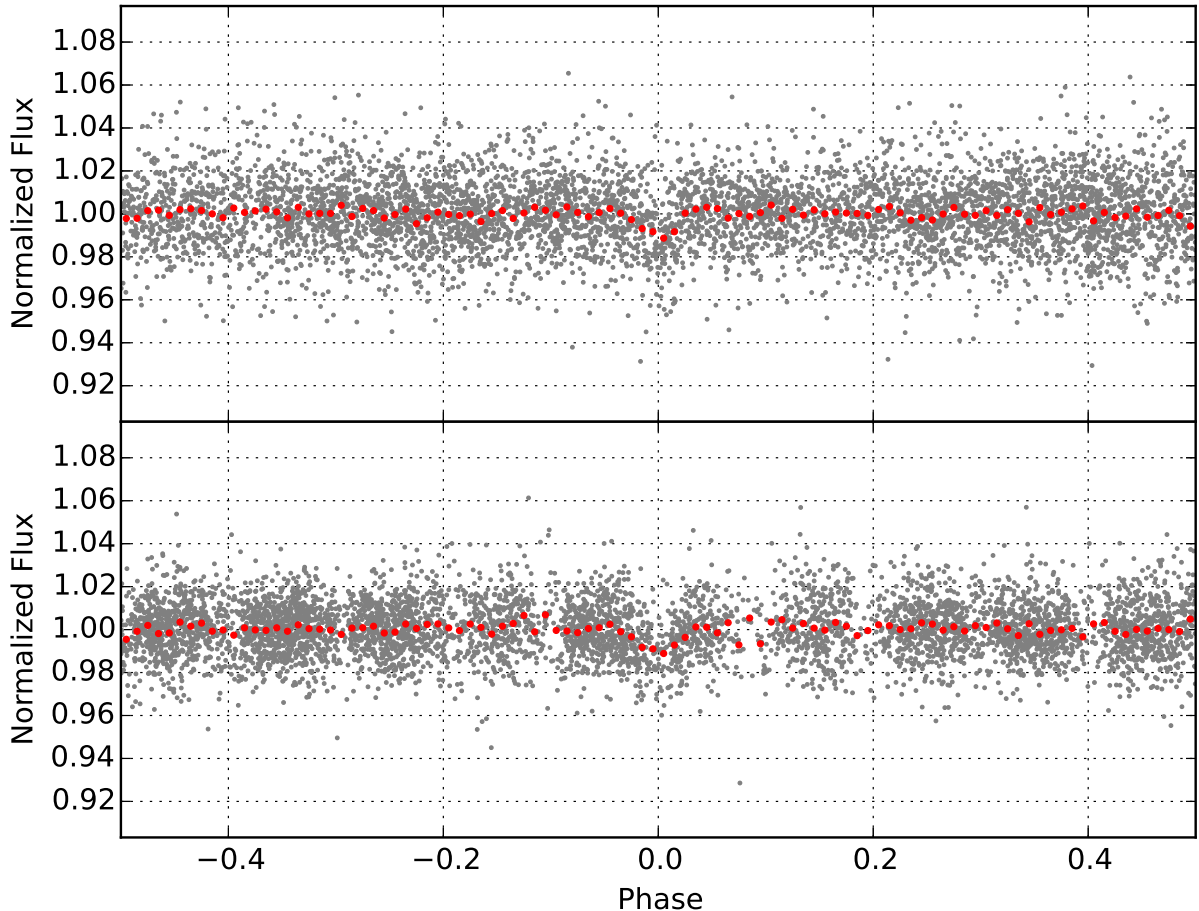,clip=,width=0.9\linewidth}
\caption{Discovery light curve of KELT-14b (Top) and KELT-15b (Bottom) from the KELT-South telescope. The light curves are phase-folded to the discovery periods of P = 1.7100596 and 3.329442 days respectively; the red points show the light curve binned in phase using a bin size of 0.01. }
\label{figure:DiscoveryLC}
\end{figure*}



\begin{figure}
\includegraphics[width=1\linewidth]{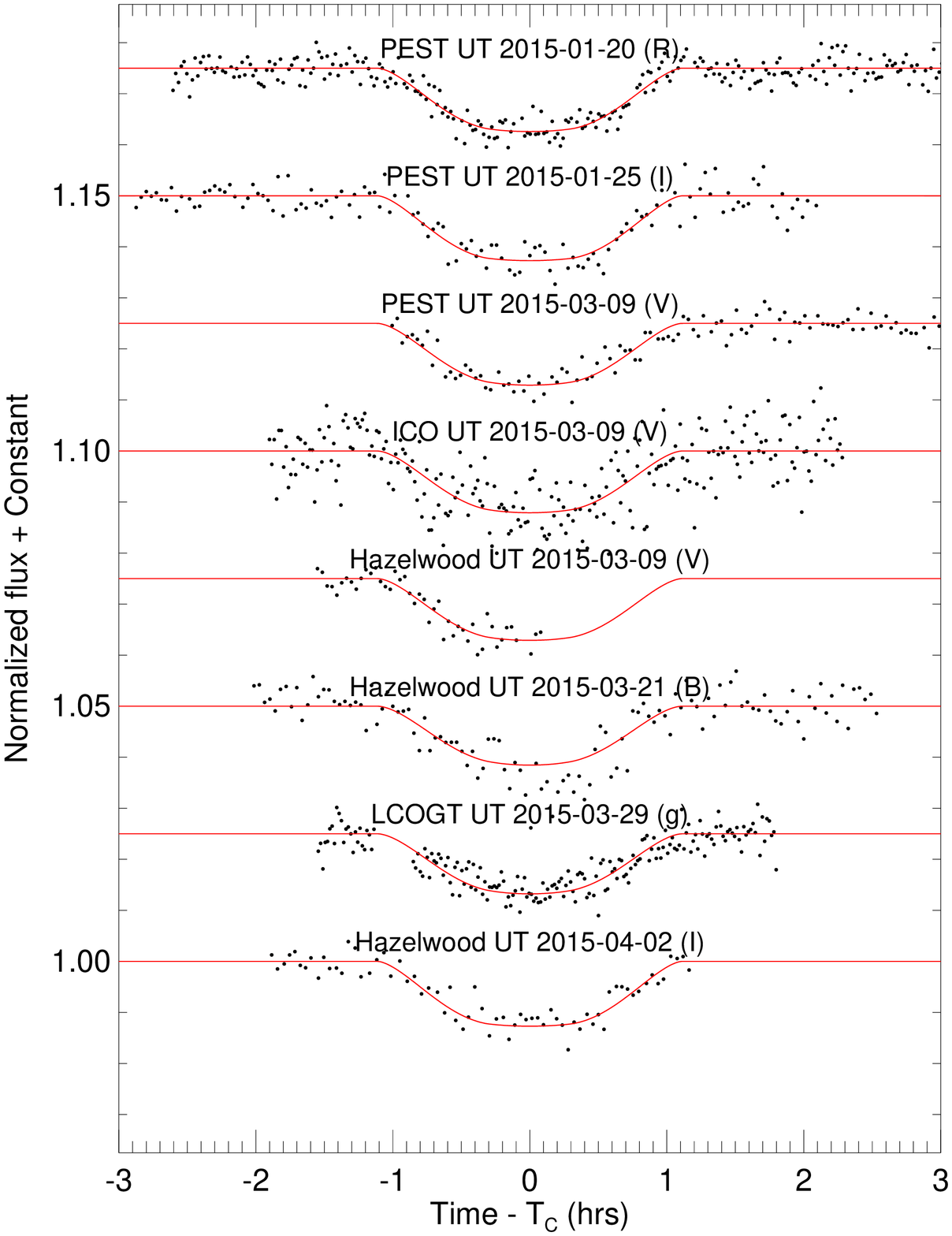}
   \vspace{-.2in}

\includegraphics[width=1\linewidth]{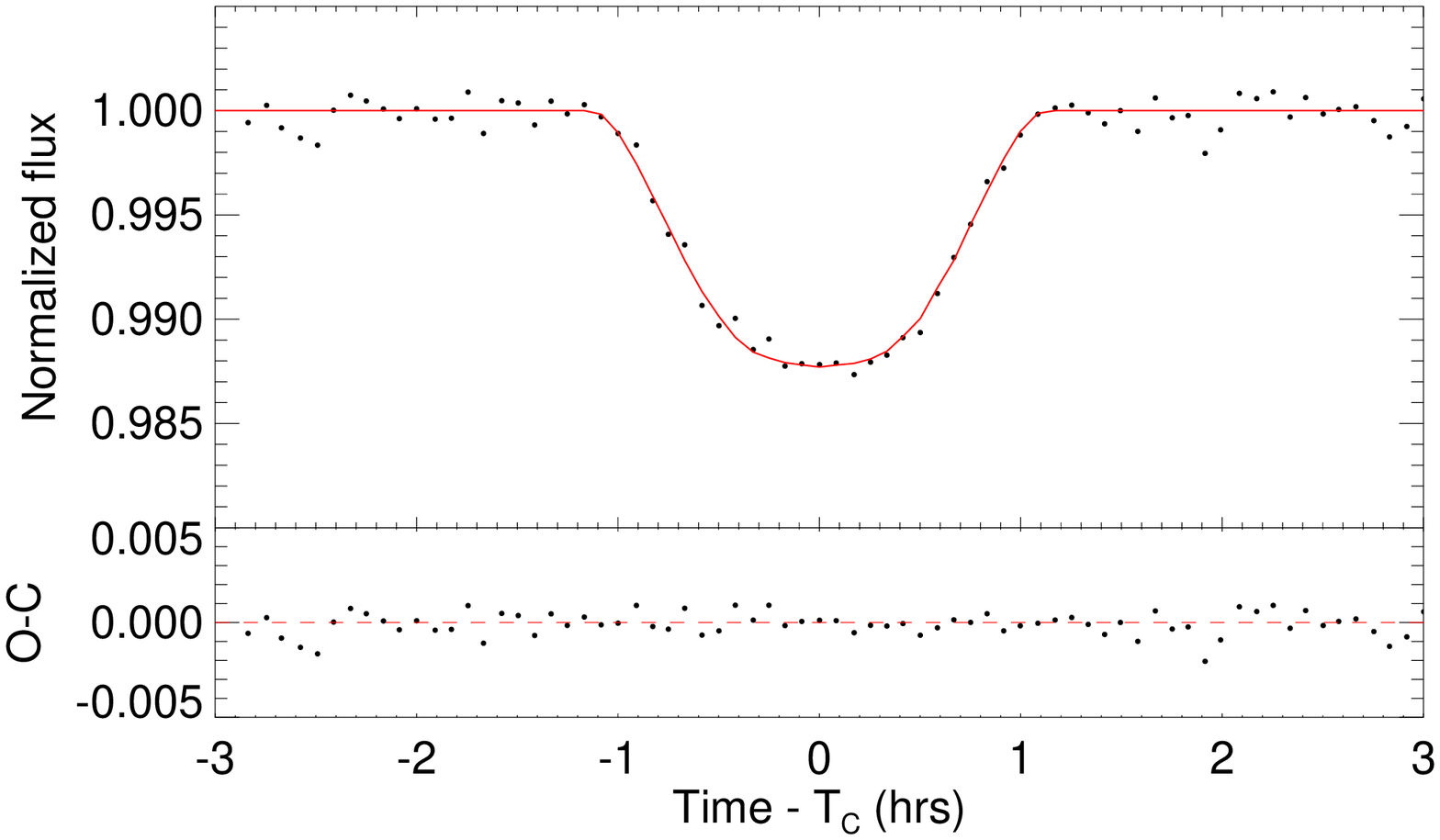}
\caption{(Top) The follow-up photometry of KELT-14b from the KELT follow-up network. The red line is the best model for each follow-up lightcurve. (Bottom) The individual follow-up lightcurves combined and binned in 5 minute intervals. This combined and binned plot represents the true nature of the transit. The combined and binned light curve is for display and is not used in the analysis. The red line represents the combined and binned individual models (red) of each follow-up observation. }
\label{fig:K14Followup} 
\end{figure}

\section{\bf{Discovery and Follow-Up Observations}}
\subsection{KELT-South}

KELT-14 and KELT-15 are located in the KELT-South field 34, which is centered at J2000 $\alpha =$ 08$^{h}$ 16$^{m}$ 12$^{s}$ $\delta =$ -54$\degr$ 00$\arcmin$ 00$\arcsec$. Field 34 was monitored in two separate campaigns: First from UT 2010 January 03 to UT 2010 February 19 as part of the KELT-South commissioning campaign, and then again from UT 2012 September 16 to UT 2014 June 14, acquiring a total of $\sim$5780 images after post-processing and removal of bad images. Following the strategy described in \citet{Kuhn:2015}, we reduced the raw images, extracted the light curves, and searched for transit candidates. Two stars emerged as top candidates from this process: KS34C030815 (TYC 7638-981-1, GSC 07638-00981, 2MASS J07131235-4224350) located at $\alpha =$ 07$^{h}$ 13$^{m}$ 12$\fs$347 $\delta =$ -42$\degr$ 24$\arcmin$ 35$\arcsec$17 J2000, hereafter as KELT-14, and KS34C034621 (TYC 8146-86-1,GSC 08146-00086, 2MASS J07493960-5207136) located at $\alpha =$ 07$^{h}$ 49$^{m}$ 39$\fs$606 $\delta =$ -52$\degr$ 07$\arcmin$ 13$\arcsec$58 J2000, designated as KELT-15 (see Figure \ref{figure:DiscoveryLC}). The host star properties for both targets are listed in Table \ref{tbl:Host_Lit_Props}. We used the box-fitting least squares (BLS) algorithm \citep{Kovacs:2002, Hartman:2012} to select these candidates, and the BLS selection criteria and values for both are shown in Table \ref{tbl:BLS_Selection_Criteria}.

\begin{table*}
\centering
\caption{Stellar Properties of KELT-14 and KELT-15 obtained from the literature.}
\label{tbl:Host_Lit_Props}
\begin{tabular}{llcccl}
   \hline
  \hline
\hline
  Parameter & Description & KELT-14 Value & KELT-15 Value & Source & Reference(s) \\
 \hline
			& 					&  TYC 7638-981-1	& 	TYC 8146-86-1	&			\\
			& 					& GSC 07638-00981		& 	GSC 08146-00086	&			\\
			&					&	2MASS J07131235-4224350 			&	2MASS J07493960-5207136	&			\\
						&					&				&		&			\\

$\alpha_{J2000}$	&Right Ascension (RA)& 07:13:12.347	& 07:49:39.606		& Tycho-2	& \citet{Hog:2000}	\\
$\delta_{J2000}$	&Declination (Dec)	& -42:24:35.17 		&-52:07:13.58 & Tycho-2	& \citet{Hog:2000}	\\
						&					&				&		&			\\
NUV    &        &    17.06$\pm$ 0.1   & N/A &   GALEX\\

B$_T$			&Tycho B$_T$ magnitude& 11.963 	&11.889 & Tycho-2	& \citet{Hog:2000}	\\
V$_T$			&Tycho V$_T$ magnitude& 11.088 		&11.440& Tycho-2	& \citet{Hog:2000}	\\
			&					&				&		&	&		\\
Johnson V		&APASS magnitude& 10.948 $\pm$ 0.05&    11.189$\pm$ 0.05		& APASS 	& \citet{Henden:2015}	\\
Johnson B		&APASS magnitude& 11.64 $\pm$ 0.05&	11.745$\pm$ 0.05	& APASS 	& \citet{Henden:2015}	\\
Sloan g'		&APASS magnitude& 11.247 $\pm$ 0.051&	11.438 $\pm$ 0.03	& APASS 	& \citet{Henden:2015}	\\
Sloan r'		&APASS magnitude& 10.733 $\pm$ 0.053&   11.048 $\pm$ 0.03		& APASS 	& \citet{Henden:2015}	\\
Sloan i'		&APASS magnitude& 10.631 $\pm$ 0.05&    10.935 $\pm$ 0.05		& APASS 	& \citet{Henden:2015}	\\
			&					&				&	&	&			\\
J			&2MASS magnitude& 9.808 $\pm$ 0.024		& 10.205 $\pm$ 0.024 &2MASS 	& \citet{Cutri:2003}	\\
H			&2MASS magnitude& 9.487	 $\pm$ 0.024		& 9.919	 $\pm$ 0.023& 2MASS 	& \citet{Cutri:2003}	\\
K			&2MASS magnitude& 9.424 $\pm$ 0.023		& 9.854 $\pm$ 0.025& 2MASS 	& \citet{Cutri:2003}	\\
			&					&				&		&			\\
\textit{WISE1}		&WISE passband& 9.369 $\pm$ 0.023		&9.775 $\pm$ 0.023 & WISE 		&\citet{Cutri:2012}	\\
\textit{WISE2}		&WISE passband& 9.414 $\pm$ 0.021		& 9.805 $\pm$ 0.020& WISE 		& \citet{Cutri:2012}\\
\textit{WISE3}		&WISE passband& 9.339 $\pm$ 0.026		& 9.919 $\pm$ 0.048& WISE 		& \citet{Cutri:2012}	\\
\textit{WISE4}		&WISE passband& 9.442 $\pm$ 0.495	& $<$\mynotes{9.580}				& WISE 		& \citet{Cutri:2012}	\\
			&					&				&		&			\\
$\mu_{\alpha}$		& Proper Motion in RA (mas yr$^{-1}$)	& -13.9 $\pm$ 2.2		&  	-3.4 $\pm$ 2.3	& NOMAD		& \citet{Zacharias:2004} \\
$\mu_{\delta}$		& Proper Motion in DEC (mas yr$^{-1}$)	&  -1.3 $\pm$ 2.0	&  -2.0 $\pm$ 2.9	& NOMAD		& \citet{Zacharias:2004} \\
			&					&				&		&			\\
U$^{*}$ & Space motion (\kms) &   -4.6 $\pm$ 1.9 & 7.8 $\pm$ 3.8 & & This work \\
V & Space motion (\kms) &   -14.6 $\pm$ 0.9 & 2.6 $\pm$ 0.8 & & This work \\
W & Space motion (\kms) &   -14.0 $\pm$ 2.3 & -1.5 $\pm$ 3.3  & & This work \\
Distance & Estimated Distance (pc) & 201$\pm$19 & 291$\pm$30 & & This work \\
RV & Absolute RV (\kms) &  34.62 $\pm$ 0.13   & 12.20 $\pm$ 0.11 & & This work \\
$v\sin{i_*}$&  Stellar Rotational Velocity (\kms)       &7.7$\pm$0.4 &  7.6$\pm$0.4 & & This work \\
\hline
\hline
\end{tabular}
 \footnotesize \textbf{\textsc{NOTES}} \\
 \footnotesize Red value correspond to upper limits (S/N $<$ 2)

\footnotesize $^{*}$U is positive in the direction of the Galactic Center 
\end{table*}

\begin{table}
 \caption{KELT-South BLS selection criteria}
\small
\label{tbl:BLS_Selection_Criteria}
 \begin{tabular}{@{}llll}
    \hline
    BLS  &  Selection  & KELT-14b & KELT-15b\\
   Statistic &  Criteria & KS34C030815 & KS34C034621\\
    \hline
    Signal detection \dotfill & SDE $>$ 7.0 & 7.75403& 11.04677\\
       efficiency\dotfill & & & \\
   Signal to pink-noise\dotfill & SPN $>$ 7.0 & 8.26194&9.78164\\
    Transit depth\dotfill & $\delta <$ 0.05 & 0.01072&0.00841\\
    $\chi^2$ ratio\dotfill &  $\displaystyle\frac{\Delta\chi^2}{\Delta\chi^2_{-}} >$ 1.5 & 2.16 &2.56 \\
    Duty cycle\dotfill & q $<$ 0.1 &0.03333&0.04667\\
    \hline
 \end{tabular}
\end{table}  

\subsection{Photometric Follow-up}
\begin{table*}
 \centering
 \caption{Photometric follow-up observations and the detrending parameters found by AIJ for the global fit.}
 \label{tbl:detrending_parameters}
 \begin{tabular}{lllllllll}
    \hline
    \hline
    Target & Observatory & Date (UT) & Filter & FOV & Pixel Scale  & Exposure (s) & FWHM & Detrending parameters for global fit \\
    \hline
    KELT-14b & PEST & UT 2015 January 20 & $R$ & 31 $\arcmin$ $\times$ 21 $\arcmin$&  1.2$\arcsec$ & 60 & 6.04 & airmass, \textit{y} coordinates \\
    KELT-14b & PEST & UT 2015 January 25  & $I$ &  31 $\arcmin$ $\times$ 21 $\arcmin$&  1.2$\arcsec$ & 120 &7.48 &airmass, \textit{y} coordinates \\
    KELT-14b & PEST & UT 2015 March 09  & $V$ &  31 $\arcmin$ $\times$ 21 $\arcmin$&  1.2$\arcsec$ &120 &5.56 &airmass \\
    KELT-14b & Adelaide & UT 2015 March 09  & $V$ & 16.6$\arcmin$ $\times$ 12.3$\arcmin$ & 0.62$\arcsec$ &60 &10.48 &airmass, total counts \\
    KELT-14b & Hazelwood & UT 2015 March 09  & $V$ & 18$\arcmin$ $\times$ 12$\arcmin$ & 0.73$\arcsec$ & 120&6.10 &airmass \\
    KELT-14b & Hazelwood & UT 2015 March 21 & $B$ & 18$\arcmin$ $\times$ 12$\arcmin$ & 0.73$\arcsec$ &120 &6.31 &airmass\\
    KELT-14b &LCOGT & UT 2015 March 29 & $g^{\prime}$ &27$\arcmin$ $\times$ 27$\arcmin$ & 0.39$\arcsec$&39&11.24 &airmass, pixel width, total counts \\
    KELT-14b &Hazelwood & UT 2015 April 02 & $I$ & 18$\arcmin$ $\times$ 12$\arcmin$ & 0.73$\arcsec$ & 120& 7.19 &airmass \\
        \hline
    KELT-15b & Adelaide & UT 2014 December 27 & $V$ & 16.6$\arcmin$ $\times$ 12.3$\arcmin$ & 0.62$\arcsec$ &60 &9.95 &airmass, \textit{y} coordinates, total counts \\
    KELT-15b & Adelaide & UT 2015 January 06 & $R$ & 16.6$\arcmin$ $\times$ 12.3$\arcmin$ & 0.62$\arcsec$ &120 &13.8 &airmass, \textit{y} coordinates \\
    KELT-15b & PEST & UT 2015 January 16 & $I$ &  31 $\arcmin$ $\times$ 21 $\arcmin$ &  1.2$\arcsec$ &120 &6.35 &airmass, sky counts per pixel, total counts \\

    \hline
    \hline
 \end{tabular}
\begin{flushleft}
  \footnotesize \textbf{\textsc{NOTES}} \\
  \footnotesize All the follow-up photometry presented in this paper is available in machine-readable form in the online journal.
  \end{flushleft}
\end{table*}

\begin{figure}
\includegraphics[width=1\linewidth]{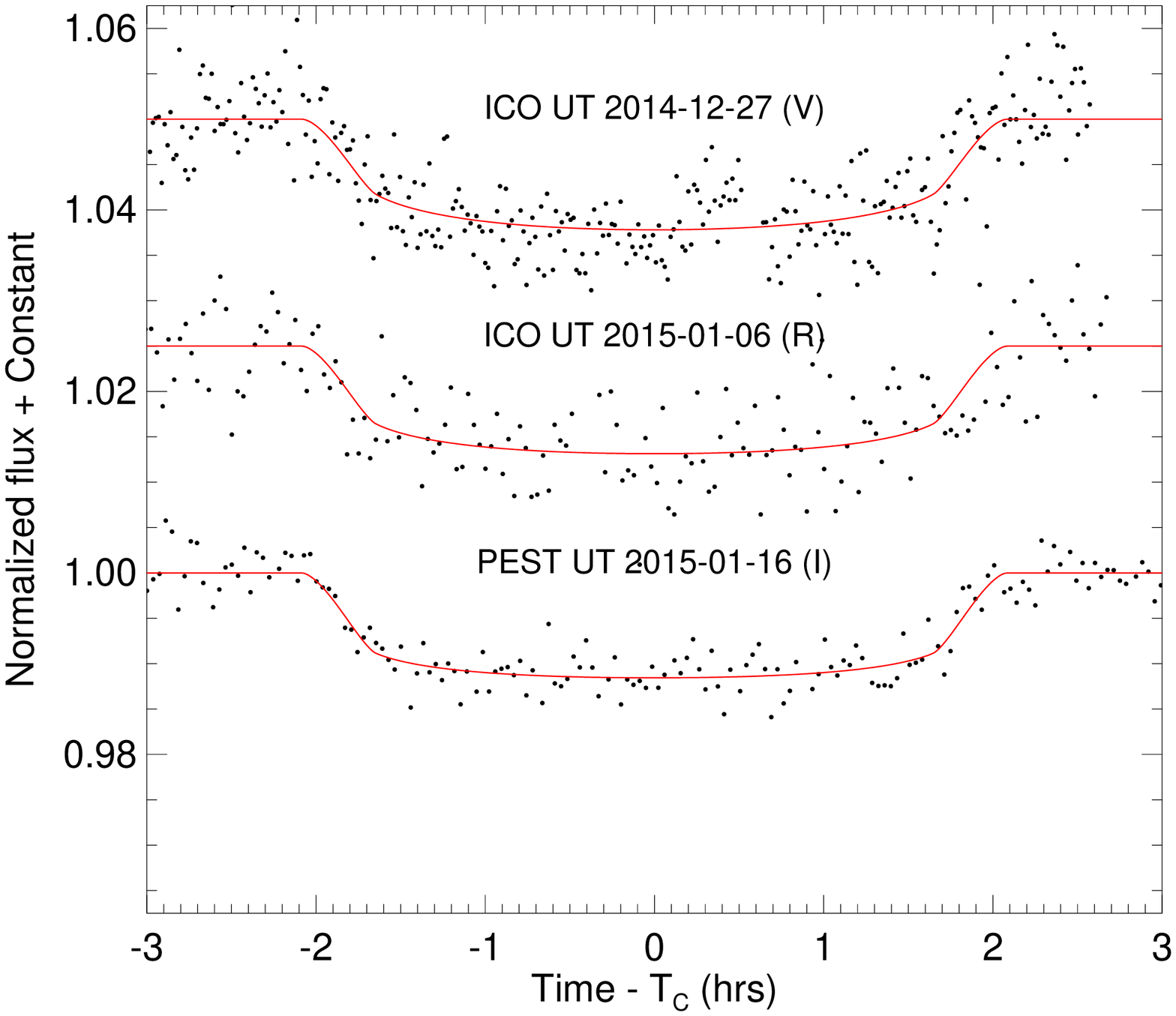}
   \vspace{-.2in}

\includegraphics[width=1\linewidth]{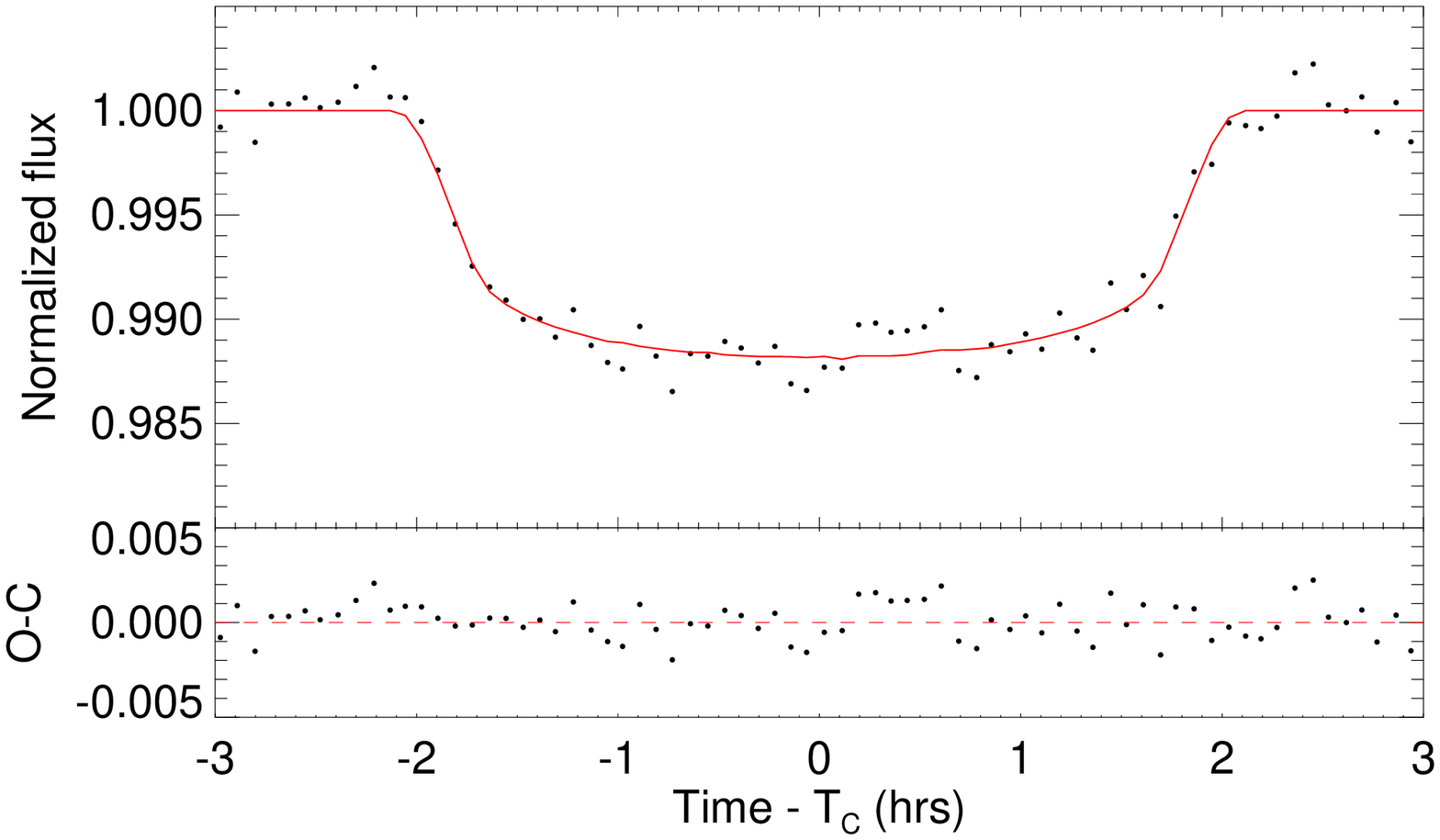}
\caption{(Top) The follow-up photometry of KELT-15b from the KELT follow-up network. The red line is the best model for each follow-up lightcurve. (Bottom) All the follow-up lightcurves combined and binned in 5 minute intervals. This best represents the true nature of the transit. The combined and binned light curve is for display and is not used in the analysis. The red line represents the combined and binned individual models (red) of each follow-up observation.}
\label{fig:K15Followup} 
\end{figure}

To precisely measure the transits of KELT-14b and KELT-15b, we obtained high-cadence, high-precision photometric follow-up using larger telescopes that cleanly resolve the hosts from their neighbors within a few arcseconds. These observations better constrain the period, depth, and duration of the transit and also rule out various false positive scenarios. To predict the transits, we use the web interface, TAPIR \citep{Jensen:2013}. For consistency, all follow-up observations were analyzed using AstroImageJ (AIJ) \citep{Collins:2013, Collins:2015}. This software also provides the best detrending parameters that are included in the global fit (see \S \ref{sec:Global_Modeling}). The follow-up photometry for KELT-14b and KELT-15b are shown in Figures \ref{fig:K14Followup} and \ref{fig:K15Followup} respectively.

\subsubsection{LCOGT}
 
We observed a nearly full transit of KELT-14b in the Sloan $g$-band on UT 2015 March 29 from a 1-m telescope in the Las Cumbres Observatory Global Telescope (LCOGT) network$\footnote{http://lcogt.net/}$ located at Cerro Tololo Inter-American Observatory (CTIO) in Chile. The LCOGT telescopes at CTIO have a 4K $\times$ 4K Sinistro detector with a 27$\arcmin$ $\times$ 27$\arcmin$ field of view and a pixel scale of 0.39$\arcsec$ per pixel. The typical FWHM of the star in this data set was 11.24 pixels. The reduced data were downloaded from the LCOGT archive and analyzed using the AstroImageJ software. In a portion of the light curve surrounding the transit ingress the target was saturated, therefore we exclude this portion of the data from the global parameter analysis in \S \ref{sec:Global_Modeling}. 

\subsubsection{PEST Observatory}
PEST (Perth Exoplanet Survey Telescope) observatory is a home observatory owned and operated by Thiam-Guan (TG) Tan. It is equipped with a 12-inch Meade LX200 SCT f/10 telescope with focal reducer yielding f/5. The camera is an SBIG ST-8XME with a filter wheel having $B$, $V$, $R$, $I$ and Clear filters. The focusing is computer controlled with an Optec TCF-Si focuser. The image scale obtained is 1.2 $\arcsec$ per pixel and a full frame image covers 31 $\arcmin$ $\times$ 21 $\arcmin$.  For images in focus the usual star FWHM achieved is about 2.5 to 3.5 pixels. The PEST observatory clock is synced on start up to the atomic clock in Bolder, CO and is resynced every 3 hours. PEST observed full transits of KELT-14b on UT 2015 January 20 ($R$) and UT 2015 January 25 ($I$), and a nearly full transit on UT 2015 March 09 ($V$). PEST observed a full transit of KELT-15b on UT 2015 January 16 ($I$).

\subsubsection{Hazelwood Observatory}
The Hazelwood Observatory is a backyard observatory with 0.32 m Planewave CDK telescope working at f/8, a SBIG ST8XME 1.5K $\times$ 1K CCD, giving a 18$\arcmin$ $\times$ 12$\arcmin$ field of view and 0.73$\arcsec$ per pixel. The camera is equipped with Clear, B, V, Rc, and Ic filters (Astrodon Interference). Typical FWHM is 2.4$\arcsec$ to 2.7$\arcsec$. The Hazelwood Observatory, operated by Chris Stockdale in Victoria, Australia, obtained an ingress of KELT-14b in $V$-band on UT 2015 March 09, a full transit in the $B$-band on UT 2015 March 21 and a full transit in $I$-band on UT 2015 April 02. The observatory computer clock is synchronised at the start of each observing session and then every 15 minutes using NTP protocol to time.nist.gov. ACP, ACP Scheduler and MaximDL are used to acquire the images. The camera shutter latency (0.5s) is allowed for within MaximDL and the adjusted exposure time is recorded within the FITS header. Experience with another project has shown that the exposure start time is recorded in the FITS header to within one second of the actual exposure start time.

\subsubsection{Adelaide Observations}
The Adelaide Observatory, owned and operated by Ivan Curtis is located in Adelaide, Australia (labeled ``ICO" in the figures). The observatory is equipped with a 9.25-in Celestron SCT telescope with an Antares 0.63x focal reducer yielding an overall focal ratio of f/6.3. The camera is an Atik 320e, which uses a cooled Sony ICX274 CCD of $1620 \times 1220$ pixels. The field of view is 16.6$\arcmin$ $\times$ 12.3$\arcmin$ with a pixel scale of 0.62$\arcsec$ per pixel and a typical FWHM around 2.5 to 3.1 $\arcsec$. The observatory's computer clock is synced with an internet time server before each observation session and has an overall timing uncertainty of a few seconds. The Adelaide Observatory observed a full transit of KELT-14b on UT 2015 March 09 ($V$) and full transits of KELT-15b on UT 2014 December 27 ($R$) and UT 2015 January 06 ($R$).

\subsection{Spectroscopic Follow-up}
\label{sec:Spectroscopy}

\begin{table*}
 \centering
 \caption{Spectroscopic follow-up observations}
 \label{tbl:spectroscopic_parameters}
 \begin{tabular}{lllllllll}
    \hline
    \hline
Target & Telescope/Instrument & Date Range & Type of Observation & Resolution & Wavelength Range & Mean S/N & Epochs\\
    \hline
KELT-14 & ANU 2.3/WiFes  &  UT 2015 Feb 02 & Reconnaissance & R$\sim3000$ & $3500-6000$\AA & 75 & 1\\
KELT-14 & ANU 2.3/WiFes  & UT 2015 Feb 02 -- UT 2015 Feb 04& Reconnaissance & R$\sim7000$ & $5200-7000$\AA & 85 & 3\\
KELT-15 & ANU 2.3/WiFes  &  UT 2014 Dec 29 & Reconnaissance & R$\sim3000$ & $3500-6000$\AA & 110 & 1\\
KELT-15 & ANU 2.3/WiFes  &  UT 2014 Dec 29 -- UT 2015 Jan 02& Reconnaissance & R$\sim7000$ & $5200-7000$\AA & 80 & 3\\
KELT-14 & AAT/CYCLOPS2  & UT 2015 Feb 26 -- UT 2015 May 13 & High Resolution & R$\sim70,000$ & $4550-7350$\AA & 41.6 & 15\\
KELT-15 & AAT/CYCLOPS2  &UT 2015 Feb 27 -- UT 2015 May 15 & High Resolution & R$\sim70,000$ & $4550-7350$\AA & 41.2 & 14\\
KELT-15 & Euler/CORALIE  & UT 2015 Sep 04 -- UT 2015 Sep 13  & High Resolution & R$\sim60,000$ & $3900-6800$\AA & 28.25 & 5\\
   \hline
    \hline
 \end{tabular}
\end{table*}

\subsubsection{Reconnaissance Spectroscopy}
\label{sec:K14Spectroscopy}

Since many astrophysical phenomena can create photometric signals that mimic planetary transits, it is important to follow up all candidates carefully to eliminate false positives. After identifying the targets as planet candidates from the KELT photometry, a first stage of spectroscopic reconnaissance was done using the WiFeS spectrograph mounted on the 2.3m ANU telescope at Siding Spring Observatory \citep{Dopita:2007}. This instrument is an optical dual-beam, image-slicing integral-field spectrograph. The full WiFeS observing strategy and reduction procedure is described in \citet{Bayliss:2013}. 

\begin{table}
 \centering
 \caption{KELT-14 radial velocity observations with CYCLOPS2.}
 \begin{tabular}{cccc}
 \hline
 \hline
  $\bjdtdb$ & RV & RV error& Instrument\\
   & (\ms)&  (\ms) &\\
 \hline
2457079.939623842 & 34621.30 & 16.70& CYCLOPS2 \\
2457079.991772522 & 34658.60 & 8.20& CYCLOPS2 \\
2457080.950428010 & 34456.10 & 5.20& CYCLOPS2 \\
2457081.937382183 & 34725.20 & 5.50& CYCLOPS2 \\
2457083.075531623 & 34431.30 & 6.00& CYCLOPS2 \\
2457148.892669835 & 34776.70 & 9.80& CYCLOPS2 \\
2457148.924568460 & 34792.00 & 12.60& CYCLOPS2 \\
2457149.929456027 & 34505.50 & 19.10& CYCLOPS2 \\
2457150.967262368 & 34528.10 & 13.10& CYCLOPS2 \\
2457151.873724511 & 34733.40 & 14.30& CYCLOPS2 \\
2457153.886260897 & 34794.20 & 13.20& CYCLOPS2 \\
2457153.916879608 & 34729.70 & 14.90& CYCLOPS2 \\
2457154.898109197 & 34468.40 & 8.30& CYCLOPS2 \\
2457155.867229521 & 34533.50 & 112.90& CYCLOPS2 \\
2457155.900058155 & 34560.60 & 111.50& CYCLOPS2 \\
\hline
 \hline
\end{tabular}
 \label{tbl:K14AAT_RV}
\begin{flushleft}
  \footnotesize \textbf{\textsc{NOTES}} \\
  \footnotesize This table is available in its entirety in a machine-readable form in the online journal.
  \end{flushleft}
\end{table}

\begin{table}
 \centering
 \caption{KELT-15 radial velocity observations with CYCLOPS2 and CORALIE.}
 \begin{tabular}{cccc}
 \hline
 \hline
  $\bjdtdb$ & RV & RV error& Instrument\\
   & (\ms)&  (\ms) & \\
 \hline
2457081.094367965 & 12320.4 & 10.8 & CYCLOPS2\\
2457083.091453823 & 12105.7 & 17.6 & CYCLOPS2\\
2457148.910598987 & 12074.4 & 16.3 & CYCLOPS2\\
2457148.942507928 & 12247.1 & 16.6 & CYCLOPS2 \\
2457149.947425124 & 12072.9 & 25.1 & CYCLOPS2 \\
2457150.985251112 & 12191.4 & 17.6 & CYCLOPS2\\
2457151.891071429 & 12281.0 & 15.0 & CYCLOPS2\\
2457151.953179348 & 12291.2 & 12.7 & CYCLOPS2\\
2457153.903635089 & 12196.2 & 16.0 & CYCLOPS2\\
2457153.934254059 & 12188.1 & 19.9 & CYCLOPS2\\
2457154.912681718 & 12334.9 & 13.3 & CYCLOPS2\\
2457154.921681414 & 12354.9 & 17.5 & CYCLOPS2\\
2457155.886209486 & 12085.3 & 118.0 &CYCLOPS2 \\
2457155.918108410 & 12057.5 & 114.8 & CYCLOPS2\\
2457269.908610  & 12096.39 &  53.81 & CORALIE \\
2457272.903199 &  12125.96  & 81.47 & CORALIE\\
2457273.907330  & 12221.40 &  57.97 & CORALIE\\
2457276.897042  & 12216.76  & 44.60 & CORALIE\\
2457278.894140  & 12161.43  & 24.16 & CORALIE\\
\hline
 \hline
\end{tabular}
 \label{tbl:K15AAT_RV}
\begin{flushleft}
  \footnotesize \textbf{\textsc{NOTES}} \\
  \footnotesize This table is available in its entirety in a machine-readable form in the online journal.
  \end{flushleft}
\end{table}

First, observations of both stars were performed at low resolution (R$\sim3000$) in the 3500-6000 {\AA} range to determine their stellar type. Both KELT-14 and KELT-15 were identified with the following parameters: KELT-14 has $\rm T_{eff} = 5572 \pm 200 K$, $\rm log g_{*} = 3.5 \pm 0.4$ (cgs)  and $\rm [Fe/H] = 0.0 \pm 0.4$; KELT-15b has $\rm T_{eff} = 6221 \pm 200 K$, $\rm log g_{*} = 3.4 \pm 0.4$ (cgs) and $\rm [Fe/H] = 0.0 \pm 0.4$. The low resolution spectra provide poor precision on the $\rm log g_{*}$ and therefore, these $\rm log g_{*}$ values aren't very reliable.

Additionally, three observations for each target were performed in medium-resolution (R$\sim7000$) using the red camera arm of the WiFeS spectrograph (5500-9000 {\AA}) across the expected orbital phase based on the photometrically detected period. These observations were aimed at performing multiple radial velocity (RV) measurements of each target to detect signals higher than 5 km/s amplitude, allowing us to identify grazing binary systems or blended eclipsing binaries. The typical RV precision achieved with this instrument is around 1.5 km/s, and both targets showed no significant variations among the three measurements.

\begin{figure}
\includegraphics[width=1\linewidth]{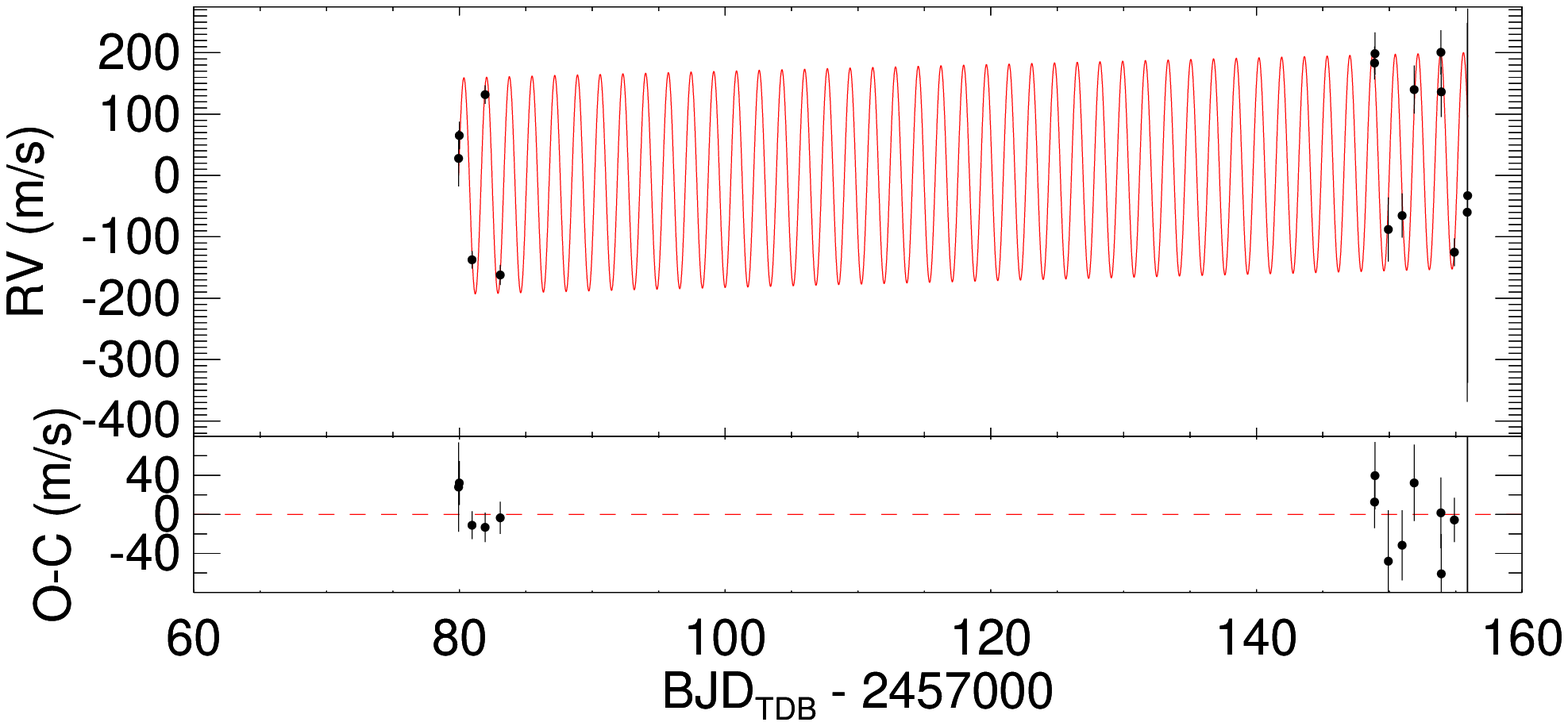}
   \vspace{-.2in}

\includegraphics[width=1\linewidth]{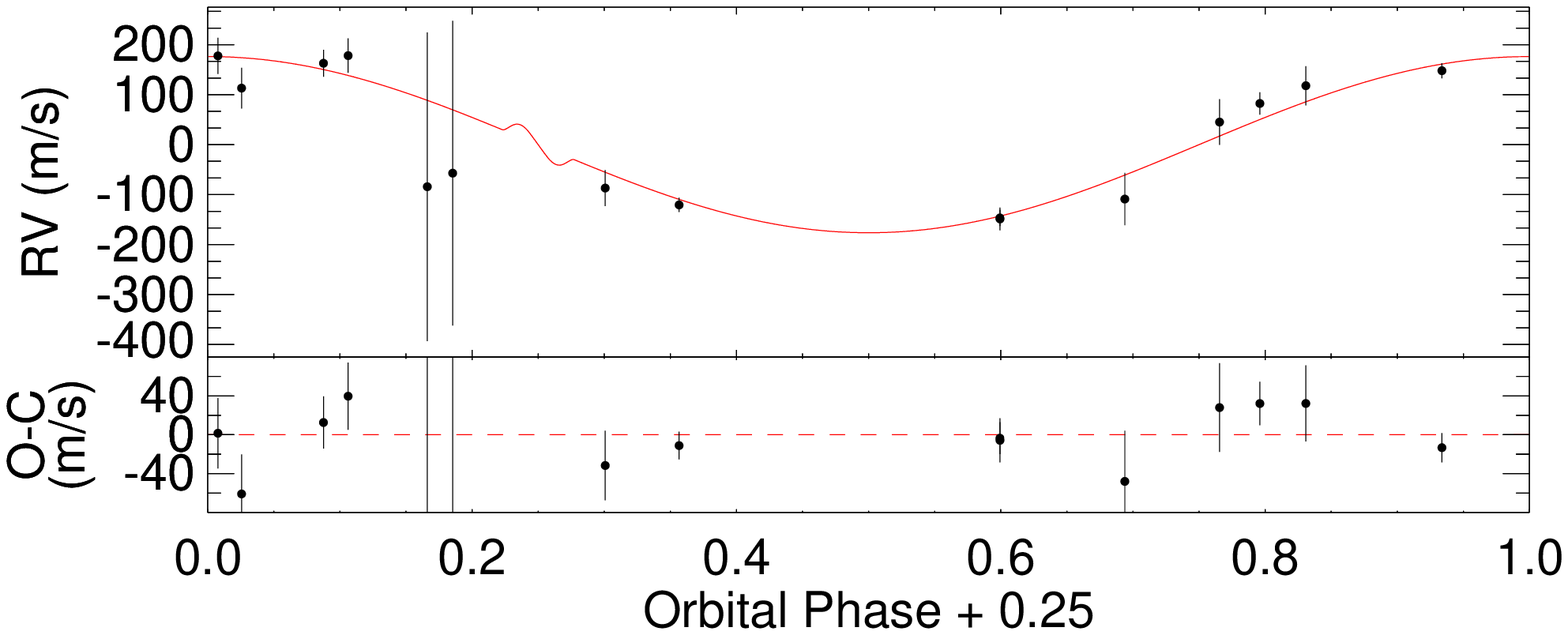}
\caption{(Top) the AAT radial velocity measurements (the median absolute RV has been subtracted off) and residuals for KELT-14. The best-fitting orbit model is shown in red. The residuals of the RV measurements to the best fitting model are shown below. (Bottom) The KELT-14 AAT measurements phase-folded to the final global fit ephemeris.}
\label{fig:K14RV} 
\end{figure}

\subsubsection{High Precision Spectroscopic Follow-up}
\label{highresRV}
To confirm the planetary nature of the companion, we obtain multi-epoch high-resolution spectroscopy. These spectra allow us to very accurately measure the radial velocity of the host star providing us with a precise measurement of the companion's mass. Also, these spectra provide a much better estimate of the stellar properties.

\subsubsection{CYCLOPS2}
\begin{figure}
\includegraphics[width=1\linewidth]{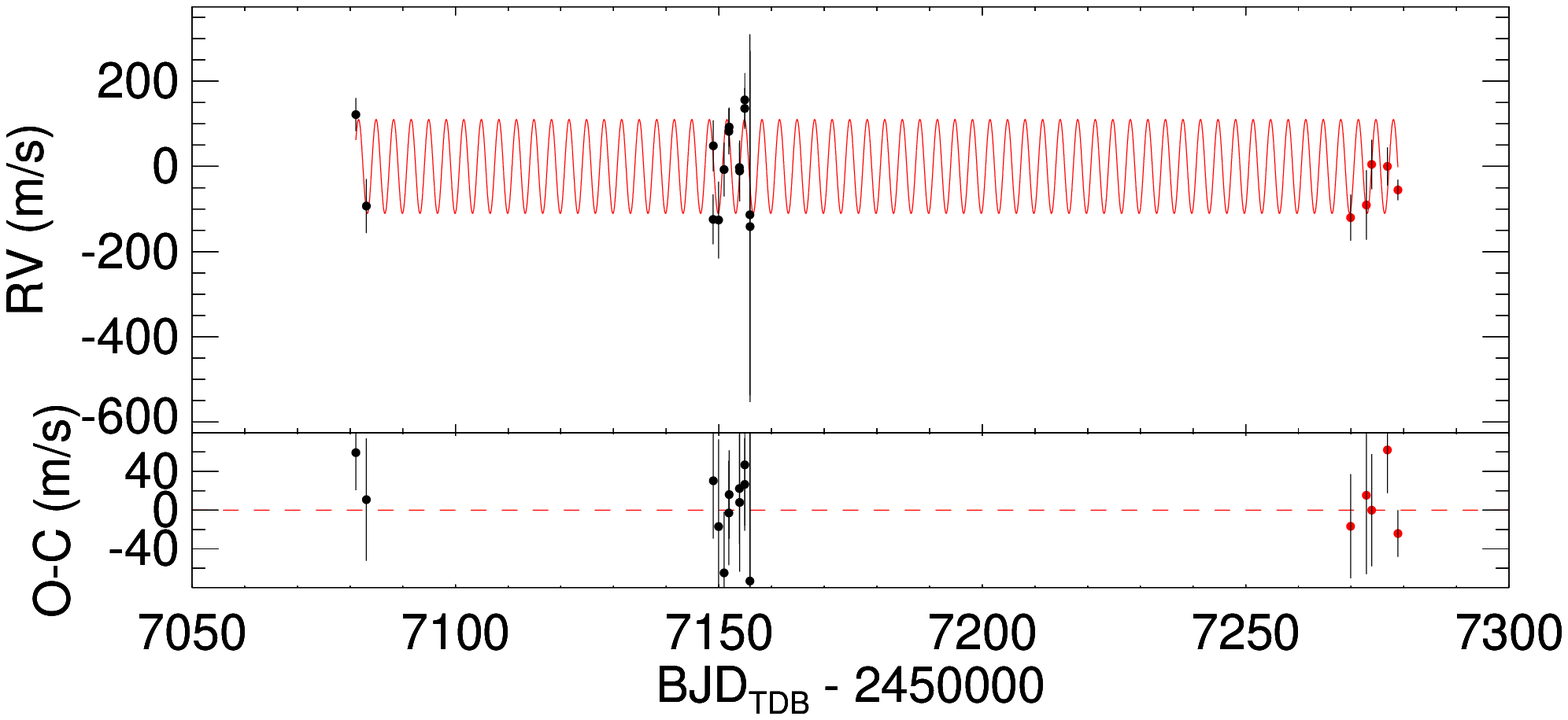}
   \vspace{-.2in}

\includegraphics[width=1\linewidth]{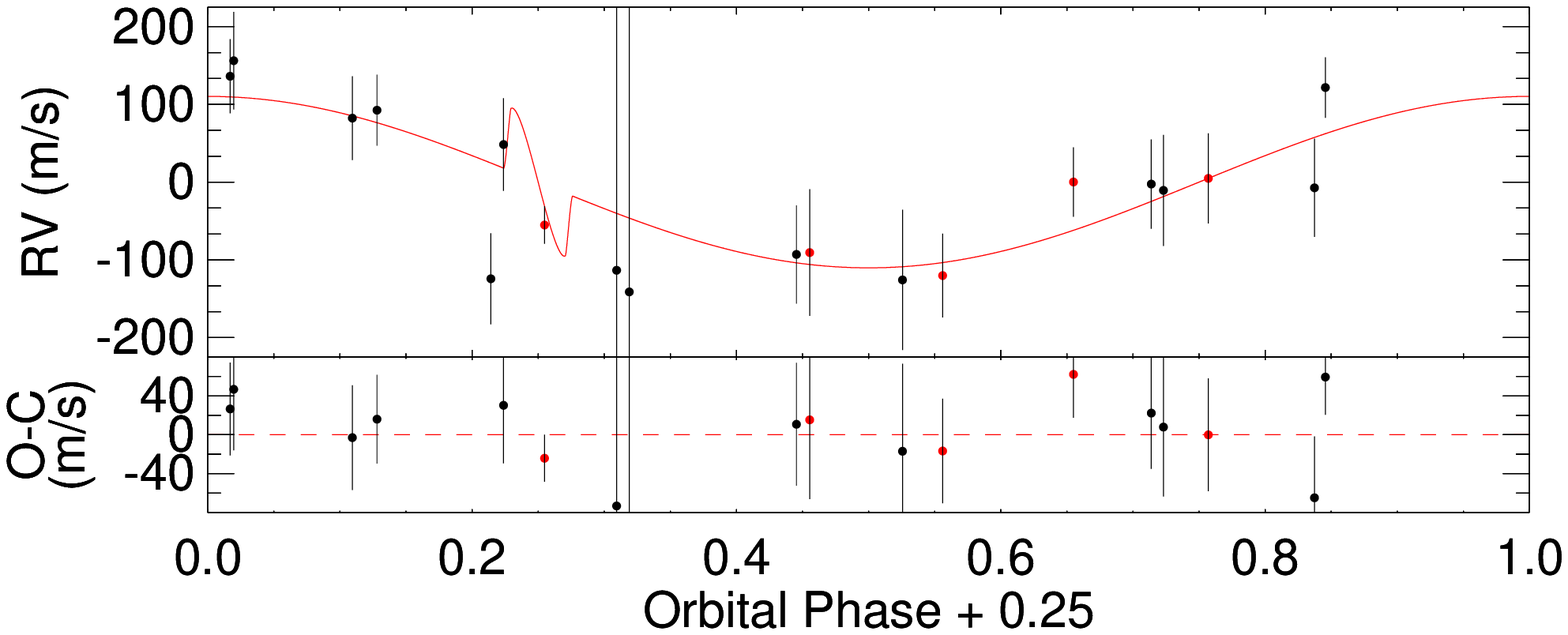}
\caption{(Top) the AAT (black) and CORALIE (red) radial velocity measurements (the median absolute RV has been subtracted off) and residuals for KELT-15. The best-fitting orbit model is shown in red. The residuals of the RV measurements to the best model are shown below. (Bottom) The KELT-15 AAT (black) and CORALIE (red) measurements phase-folded to the final global fit ephemeris.}
\label{fig:K15RV} 
\end{figure}

Spectroscopic observations of KELT-14 and KELT-15 were carried out using the CYCLOPS2 fibre feed with the UCLES spectrograph instrument on the Anglo-Australian Telescope (AAT) over two observing runs: UT 2015 February 02 - UT 2015 March 01 and UT 2015 May 6 - UT 2015 May 13 (See Figure \ref{fig:K14RV} and \ref{fig:K15RV}). The instrumental set-up and observing strategy for these observations closely follow that described in earlier CYCLOPS radial velocity papers \citep{Addison:2013,Addison:2014}. 

CYCLOPS2 is a Cassegrain fiber-based integral field unit which reformats a $\sim$2.5\arcsec\ diameter on-sky aperture into a pseudo-slit of dimensions equivalent to 0.6" wide $\times$ 14.5" long  \citep{Horton:2012}. CYCLOPS2 has 16 on-sky fibers, plus one fiber illuminated by a ThUXe lamp. Each fiber delivers a spectral resolution of $\lambda/\Delta\lambda \approx 70,000$ over 19 echelle orders in the wavelength range of $4550-7350$\AA, when used with the UCLES spectrograph in its 79 line/mm grating configuration. 

We use a ThAr calibration lamp  to illuminate all of the on-sky fibers at the beginning of observations to create a {\em reference} ThAr wavelength solution. We then use simultaneous ThUXe data from each exposure to determine low-order distortions which differentially calibrate observations through the night onto the {\em reference} ThAr solution. These reductions are carried out using custom MATLAB routines (Wright and Tinney, {\em in prep.}). Calibration precision is estimated from the scatter of fits to the simultaneous ThUXe spectral features and these are tested against velocity standards taken each night. The typical calibration precision is \textless\,10 m s$^{-1}$. This calibration error is combined with the error from a fit to the cross-correlation profile to give a final uncertainty for each observation. 

The cross-correlation profiles are obtained using a weighted cross-correlation \citep{Baranne:1996, Pepe:2002} of a stellar template produced with \textsc{synspec} \citep{Hubeny:2011}. The velocities are determined from the fit of a generalised normal distribution to the cross-correlation profiles and the errors are estimated from the Jacobian matrix for each fit. We find no correlation between the bisector spans and the measured radial velocities. This provides strong evidence against a blended eclipsing binary scenario.

\subsubsection{CORALIE}
\label{sec:coralie}

\begin{figure}
\includegraphics[angle=0,width=1\linewidth]{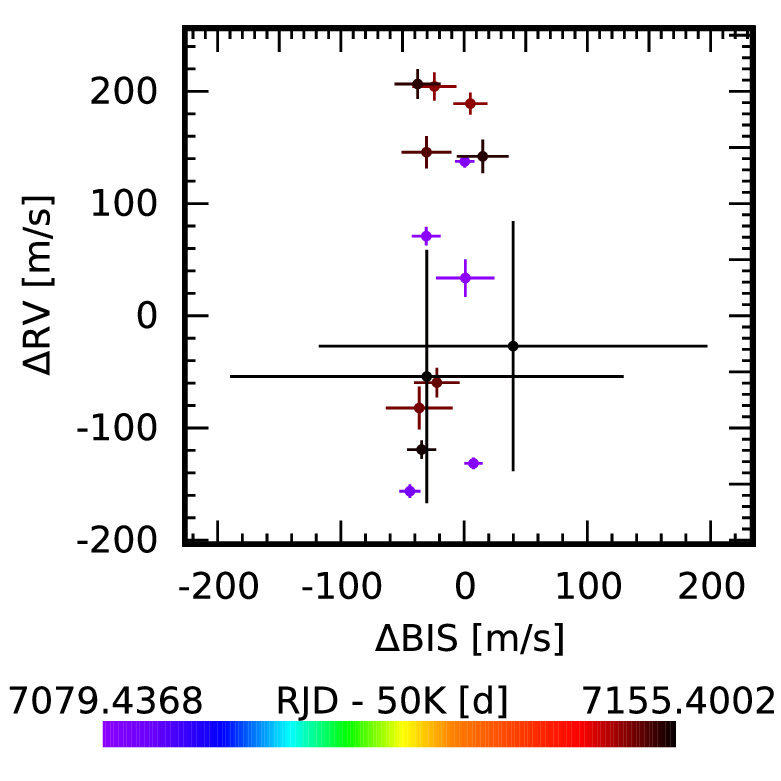}
   \vspace{-.2in}

\includegraphics[angle=0,width=1\linewidth]{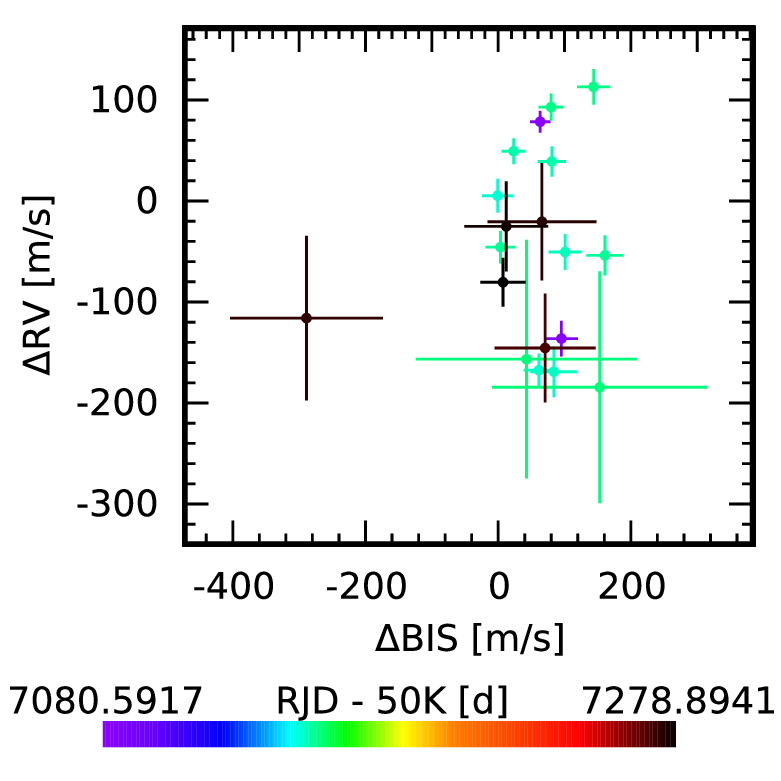}
\caption{The AAT Bisector measurements for the (Top) KELT-14 and the combined AAT and CORALIE bisector measurements for (Bottom) the KELT-15 spectra used for radial velocity measurements. We find no significant correlation between RV and the bisector spans.}
\label{fig:BIS} 
\end{figure}

CORALIE is a fibre-fed echelle spectrograph \citep{Queloz:2001} attached to the Swiss 1.2 m Leonard Euler telescope at the ESO La Silla Observatory in Chile.   It has a spectral resolution of R$\sim$60000, a wavelength range of $3900-6800$\AA, and is able to measure radial velocities of bright stars to a precision of 3 m.s$^{-1}$ or better \citep{Pepe:2002}.  In June 2015, the CORALIE spectrograph was equipped with a new Fabry-Pe\'rot-based calibration system \citep{Wildi:2011}. This system replaces the ThAr lamp for the simultaneous reference method that determines and corrects for instrumental drift occurring between the calibration and the science exposure \citep{Baranne:1996}. The data-reduction software has been adapted to take into account the new operational mode and take benefit from the higher spectral content, and hence the lower photon noise, on the drift measurement, provided by the Fabry-Pe\'rot based calibration source. We obtained spectra at five epochs of KELT-15 from UT 2015 September 02 to UT 2015 September 14. All observations were reduced and radial velocities were computed in real time using the standard CORALIE pipeline. The observations from CORALIE are consistent with the CYCLOPS2 measured radial velocities. The results are shown in Figure \ref{fig:K15RV}. We find no correlation between the bisector spans and the measured radial velocities (see Figure \ref{fig:BIS}).

\section{Analysis and Results}

\subsection{SME Stellar Analysis}\label{sec:sme}

In order to determine precise stellar parameters for KELT-14 and KELT-15, we use the available high-resolution, low S/N AAT CYCLOPS2 spectra acquired for radial velocity confirmation of the two planetary systems. For each CYCLOPS2 dataset, we took the flux weighted mean of the individual fibers, continuum normalized each spectral order, and stitched the orders into a single 1-D spectrum. We shifted each resulting spectrum to rest wavelength by accounting for barycentric motion, and median combined all observations into a single spectrum with a S/N $\sim$ 50, sufficient for detailed spectroscopic analysis.

Stellar parameters for KELT-14 and KELT-15 are determined using an implementation of Spectroscopy Made Easy (SME) \citep{Valenti:1996}. Our Monte Carlo approach to using SME for measuring stellar parameters is detailed in \citet{Kuhn:2015}. Briefly, we use a multi-trial minimization of 500 randomly selected initial parameter values, each solving for 5 free parameters: effective temperature (\teff), surface gravity ($\log{g_*}$), iron abundance (\feh), metal abundance ([m/H]), and rotational velocity of the star ($v\sin{i_*}$). We determine our final measured stellar properties by identifying the output parameters that give the optimal SME solution (i.e., the solution with the lowest $\chi^{2}$). The overall SME measurement uncertainties in the final parameters are calculated by adding in quadrature the internal error determined from the 68.3\% confidence region in the $\chi^{2}$ map, the median absolute deviation of the parameters from the 500 output SME solutions to account for the correlation between the initial guess and the final fit, and an estimate for the systematic errors in our method when compared to other common stellar spectral analysis tools \citep[see][]{Gomez:2013}.

Due to the instrument setup used for measuring high-precision radial velocities, the AAT CYCLOPS2 spectra do not include the full MgB triplet wavelength region, a pressure-broadened set of lines commonly used in spectral synthesis modeling to constrain $\log{g_*}$ \citep{Valenti:2005}. The available spectra only include one of the three strong Mg lines in this region. In order to investigate the effect of this constraint on our stellar parameters, we run two separate SME runs for both KELT-14 and KELT-15, one with $\log{g_*}$ as a free parameter and the other with $\log{g_*}$ fixed from our preliminary global fit of the photometric observations.

Our final SME spectroscopic parameters for KELT-14 are: \teff~=~5817$\pm$90~K, $\log{g_*}$~=~4.16$\pm$0.12, [m/H]~=~0.39$\pm$0.03, \feh~=~0.34$\pm$0.09 and a projected rotational velocity $v\sin{i_*}$~=7.7$\pm$0.4~\kms. Similarly, with a fixed $\log{g_*}$=4.23; \teff~=~5834$\pm$75~K, [m/H]~=~0.39$\pm$0.03, \feh~=~0.34$\pm$0.09 and $v\sin{i_*}$~=7.6$\pm$0.4~\kms. For KELT-15 we find: \teff~=~6023$\pm$61~K, $\log{g_*}$~=~3.80$\pm$0.08, [m/H]~=~0.06$\pm$0.03, \feh~=~0.05$\pm$0.03 and $v\sin{i_*}$~=11.1$\pm$0.5~\kms. With a fixed $\log{g_*}$=4.17 we find; \teff~=~6102$\pm$51~K, [m/H]~=~0.02$\pm$0.03, \feh~=~0.05$\pm$0.03 and $v\sin{i_*}$~=11.1$\pm$0.5~\kms. We constrain the macro- and microturbulent velocities to the empirically constrained relationship \citep{Gomez:2013}. However, we do allow them to change during our modelling according to the other stellar parameters. Our best fitting stellar parameters result in \textit{v}$_{\textup{mac}}$~=~4.05~\kms\space and \textit{v}$_{\textup{mic}}$~=~1.00~\kms for KELT-14, and for KELT-15 \textit{v}$_{\textup{mac}}$~=~4.37~\kms\space and \textit{v}$_{\textup{mic}}$~=~1.19~\kms.

\subsection{SED Analysis}

\begin{figure}
\includegraphics[angle=90,width=1\linewidth]{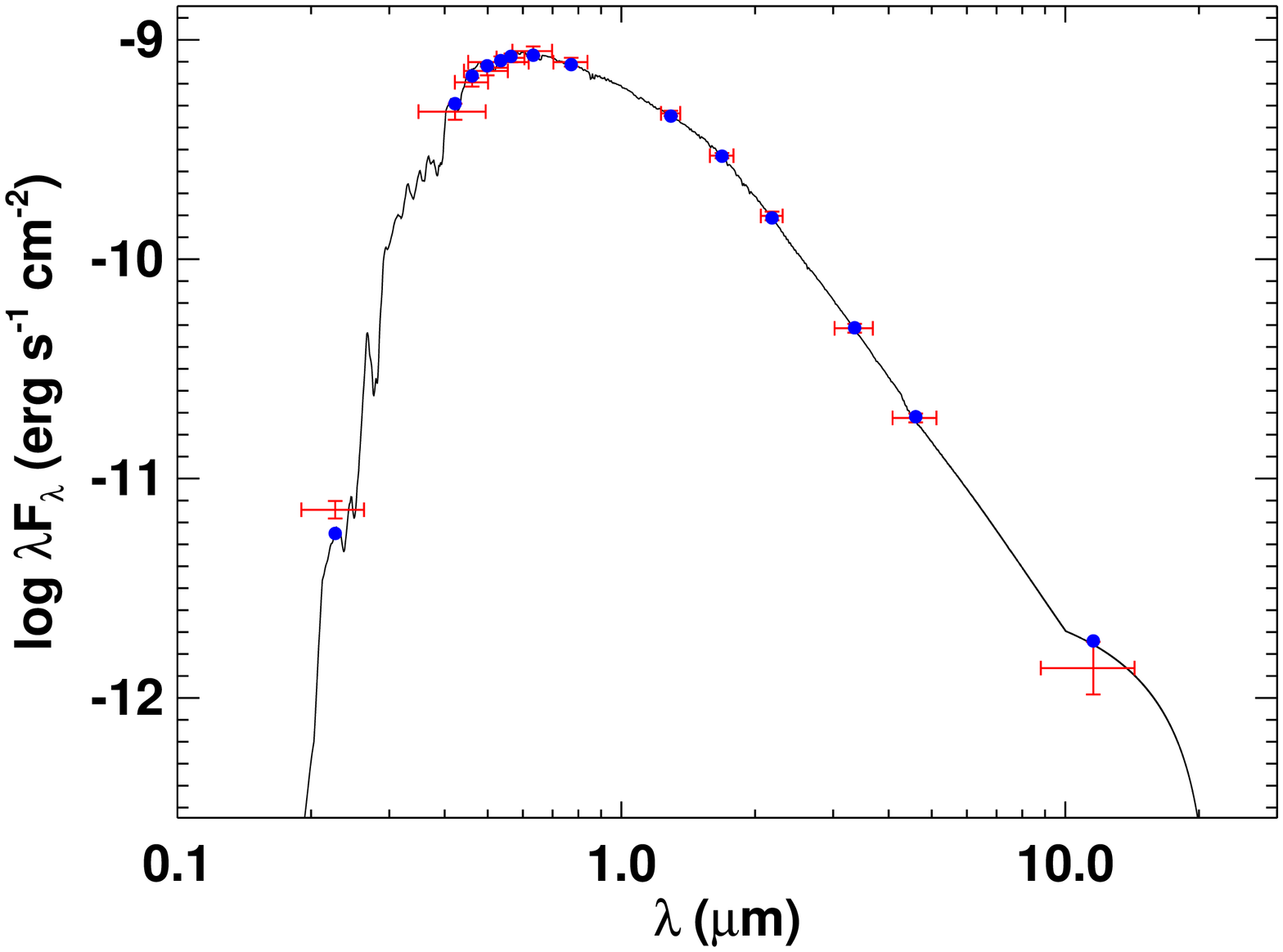}
   \vspace{-.2in}

\includegraphics[angle=90,width=1\linewidth]{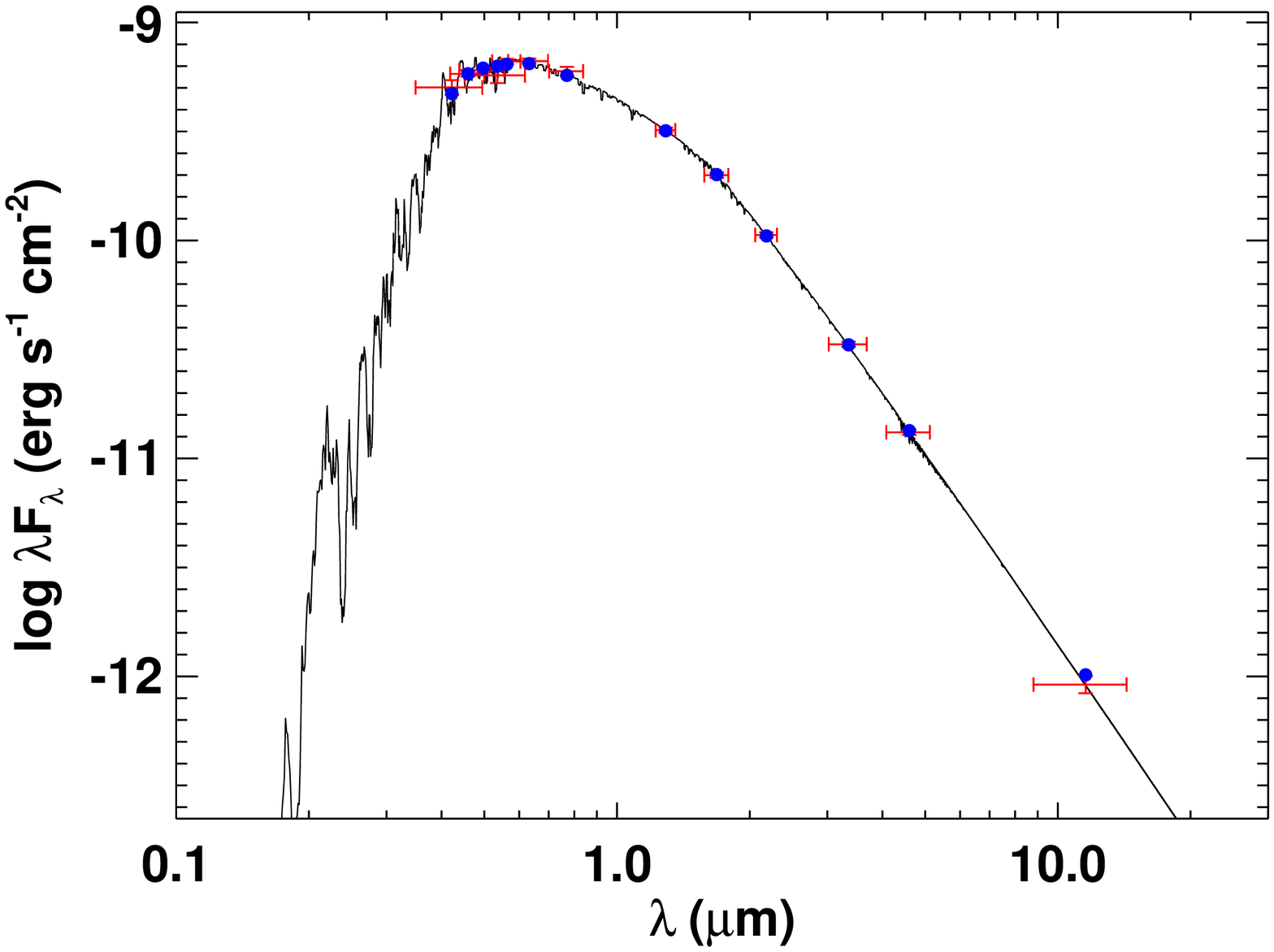}
\caption{The SED fit for (top) KELT-14 and (bottom) KELT-15. The red points show the photometric values and errors given in Table \ref{tbl:Host_Lit_Props}. The blue points are the predicted integrated fluxes at the corresponding bandpass. The black line represents the best fit stellar atmospheric model.}
\label{fig:SED} 
\end{figure}

We construct empirical spectral energy distributions (SEDs) of KELT-14 and KELT-15 using all available broadband photometry in the literature, shown in Figure \ref{fig:SED}. We use the near-UV flux from GALEX \citep{Martin:2005}, the $B_T$ and $V_T$ fluxes from the Tycho-2 catalogue, $B$, $V$, $g'$, $r'$, and $i'$ fluxes from the AAVSO APASS catalogue, NIR fluxes in the $J$, $H$, and $K_S$ bands from the 2MASS Point Source Catalogue \citep{Cutri:2003, Skrutskie:2006}, and near-and mid-infrared fluxes in the WISE passbands \citep{Wright:2010}. 

We fit these fluxes using the Kurucz atmosphere models \citep{Castelli:2004} by fixing the values of $\teff$, $\log{g_*}$ and [Fe/H] inferred from the global fit to the lightcurve and RV data as described in \S \ref{sec:Global_Modeling} and listed in Table \ref{tbl:K14AAT_RV} and Table \ref{tbl:K15AAT_RV}, and then finding the values of the visual extinction $A_V$ and distance $d$ that minimize $\chi^2$, with a maximum permitted $A_V$ based on the full line-of-sight extinction from the dust maps of \citet{Schlegel:1998} (maximum $A_V$ = 0.50 mag and 0.89 mag for KELT-14 and KELT-15, respectively). Note that while the final best SED fits below are in fact well fit with $A_V \equiv 0$, we did include $A_V$ as a free fit parameter because of the {\it a priori} likelihood of $A_V$ as large as 0.50--0.89 mag.

For KELT-14 we find A$_V = 0.1 \pm 0.1$ mag
and $d$ = 201$\pm19$ pc with the best fit model having a reduced $\chi^2 = 1.39$. For KELT-15 we find A$_V = 0.18 \pm 0.12$ and $d$ = 291$\pm30$ pc with the best fit model having a reduced $\chi^2 = 0.84$. This implies a very good quality of fit and further corroborates the final derived stellar parameters for the KELT-14 and KELT-15 host stars. We note that the quoted statistical uncertainties on $A_V$ and $d$ are likely to be underestimated because 
alternate model atmospheres would predict somewhat different SEDs and thus values of extinction and distance, but for stars of the masses and temperatures of KELT-14 and KELT-15 the systematic differences among various model atmospheres are not expected to be large. 


\subsection{Evolutionary State}
\begin{figure}
\includegraphics[angle=90,width=1\linewidth]{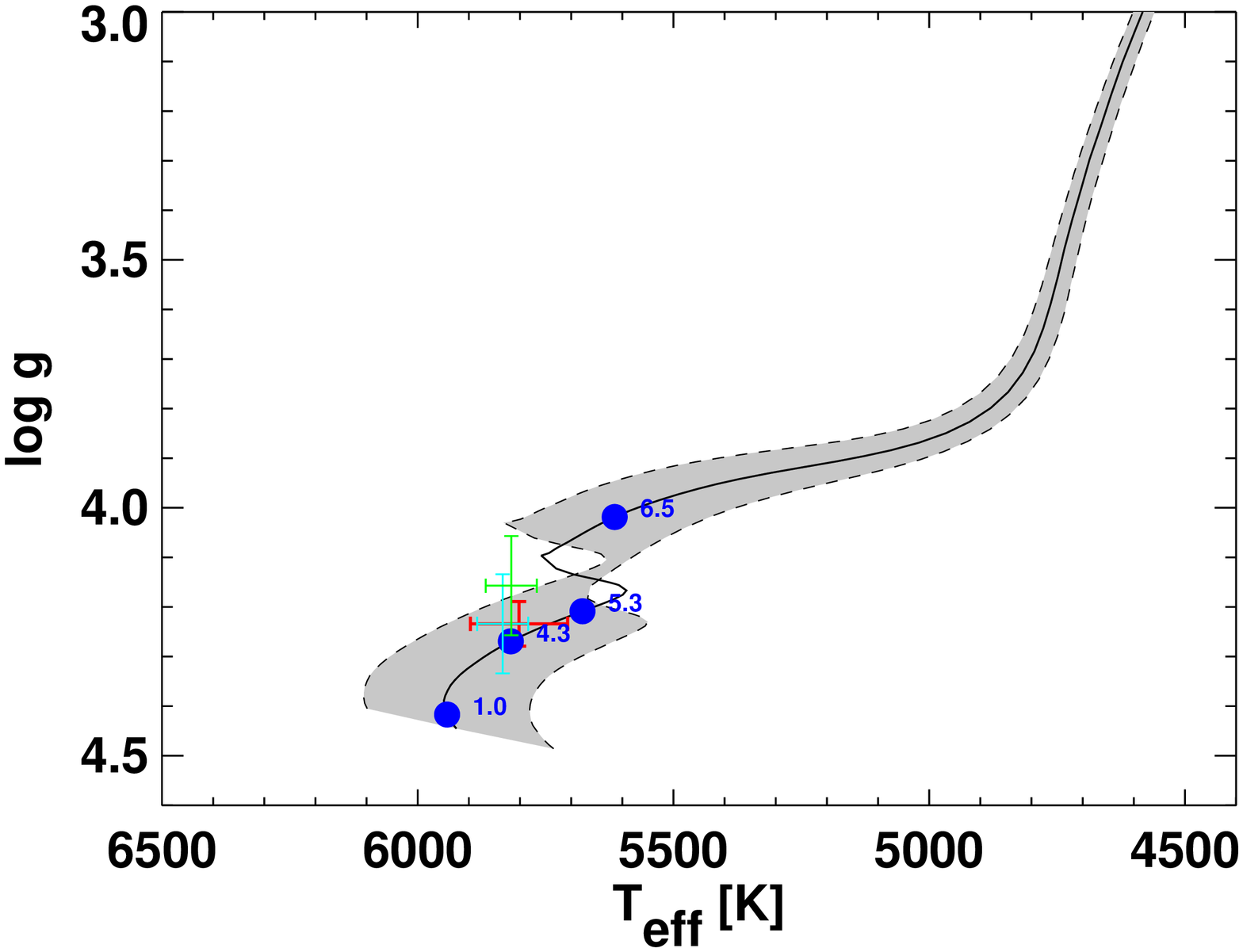}
   \vspace{-.2in}

\includegraphics[angle=90,width=1\linewidth]{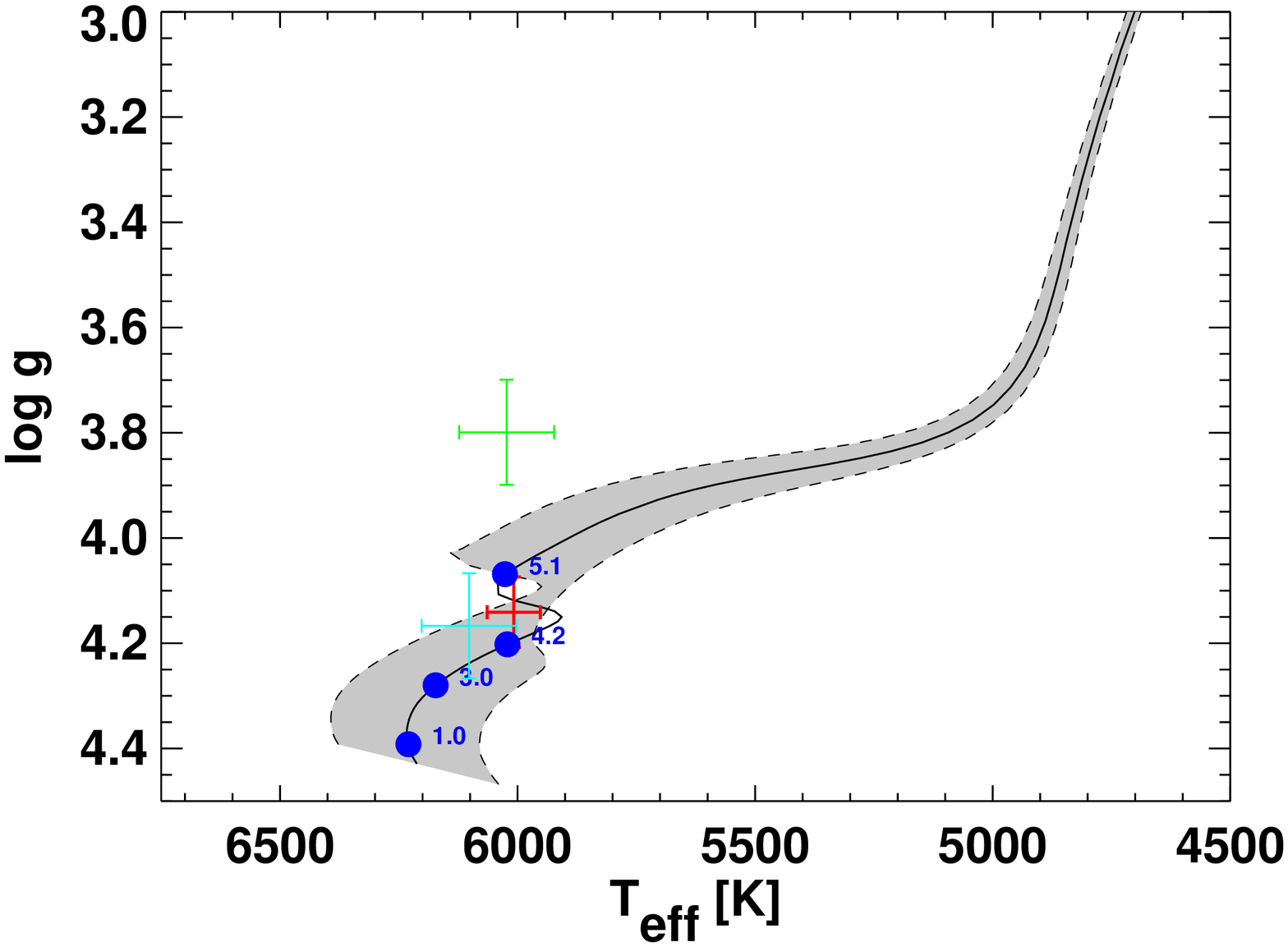}
\caption{The theoretical H-R diagrams for (top) KELT-14 and (bottom) KELT-15 using the Yonsei-Yale stellar evolution models \citep{Demarque:2004}. The $\log{g_*}$ values are in cgs units. The red cross represents the values from the final global fit. The blue cross is the position and errors of the SME analysis when $\log{g_*}$ was fixed at the initial global fit value and the green cross is when $\log{g_*}$ was not fixed. The dashed lines at the edge of the gray shaded region represent the 1$\sigma$ uncertainties on M$_{\star}$ and [Fe/H] from the global fit. The various ages along the tracks are represented by the blue points. }
\label{fig:HRD} 
\end{figure}

To better place the KELT-14 and KELT-15 systems in context, we show in Figure \ref{fig:HRD} the H-R diagrams for the two systems in the $\teff$ versus $\log{g_*}$ plane. In each case, we use the Yonsei-Yale stellar evolution model track \citep{Demarque:2004} for a star with the mass and metallicity inferred from the final global fit. Specifically, we are using the global fit where the SME determined \feh and \teff, where $\log{g_*}$ was not fixed, as priors (See Section \ref{sec:Global_Modeling}). The shaded region represents the mass and [Fe/H] fit uncertainties. The model isochrone ages are indicated as blue points, and the final best global fit $\teff$ and $\log{g_*}$ values are represented by the red error bars. For comparison, the $\teff$ and $\log{g_*}$ values determined from spectroscopy alone (without fixing $\log{g_*}$) are represented by the green error bars, while the blue error bars represent the case with $\log{g_*}$ fixed in the SME analysis (Figure \ref{fig:HRD}).

KELT-14 is a G2 type star near the main-sequence turnoff but not yet in the Hertzsprung gap, with an age of $\sim$ $5.0^{+0.3}_{-0.7}$ Gyr.  KELT-15 is a G0 type star with an age of $\sim$ $4.6^{+0.5}_{-0.4}$ Gyr, on or near the ``blue hook" just prior to the Hertzsprung gap. 
These classifications are also consistent with those reported in the catalogs of \citet{Pickles:2010} and \citet{Ammons:2006}.
Note that the observed rotational velocities of the stars (7--11 \kms; see Section \ref{sec:sme}) are consistent with the 2--15 \kms\ range observed for solar-type stars with the masses and ages of KELT-14 and KELT-15 \citep[e.g.,][]{Soderblom:1983}.

\subsection{UVW Space motion}
To better understand the place of KELT-14 and KELT-15 in the galaxy, we calculate the UVW space motion. This exercise can allow us to determine the membership and possibly the age of a star if it is associated with any known stellar groups. To calculate the UVW space motion, we combine the information presented in Table \ref{tbl:Host_Lit_Props} with the determined distance to KELT-14 and KELT-15 from the SED analysis (201$\pm$19 pc and 291$\pm$30 pc respectively). We also estimated the absolute radial velocity and error by taking the average and standard deviation of all the measured radial velocities by AAT. This gave us an estimated absolute radial velocity of 34.62 $\pm$ 0.13 \kms and 12.20 $\pm$ 0.11 \ms for KELT-14b and KELT-15b, respectively. We calculate the space motion to be U = -4.6 $\pm$ 1.9 \kms, V = -14.6 $\pm$ 0.9 \kms, W = -14.0 $\pm$ 2.3 \kms\space for KELT-14 and U = 7.8 $\pm$ 3.8 \kms, V = 2.6 $\pm$ 0.8 \kms, W = -1.5 $\pm$ 3.3 \kms\space for KELT-15 (positive U pointing toward the Galactic center). Using the peculiar velocity of the Sun with respect local standard rest (U = 8.5 \kms, V = 13.38 \kms, and W = 6.49 \kms), we have corrected for this motion in our calculations of the UVW space motion of KELT-14 and KELT-15 \citep{Coskunoglu:2011}. These space motion values give a 99\% chance that both KELT-14 and KELT-15 belong to the thin disk, according to the classification scheme of \citet{Bensby:2003}

\section{Planetary Properties}
\subsection{EXOFAST Global Fit}
\label{sec:Global_Modeling}
To perform a global fit of our photometric and spectroscopic data, we use a modified version of the IDL exoplanet fitting tool, \textsc{EXOFAST} \citep{Eastman:2013}. More detailed explanation of the global modeling is provided in \citet{Siverd:2012}. To determine a system's final parameters, simultaneous Markov Chain Monte Carlo (MCMC) analysis is performed on the AAT radial velocity measurements and the follow-up photometric observations. To constrain M$_{\star}$ and R$_{\star}$ \textsc{EXOFAST} uses either the Yonsei-Yale stellar evolution models \citep{Demarque:2004} or the empirical Torres relations \citep{Torres:2010}. Each photometric observation's raw light curve and the detrending parameters determined from the light curve are inputs for the final fit. We impose a prior on $\rm T_{eff}$ and [Fe/H] using the determined values and errors from the SME analysis of the AAT spectra. From analysis of the KELT-South and follow-up photometric observations, we set a prior on the period. For both KELT-14b and KELT-15b, we perform four global fits: 1) Using the Yonsei-Yale (YY) stellar models with eccentricity fixed at zero. 2) Using the YY stellar models with eccentricity as a free parameter. 3) Using the empirical Torres relations with eccentricity fixed at zero. 4) Using the empirical Torres relations with eccentricity as a free parameter. The results from these four global fits can be seen in Table \ref{tbl:KELT-14b} for the KELT-14 system and Table \ref{tbl:KELT-15b} for the KELT-15 system. For the parameters shown in solar or jovian units, the values for these constants are {\it G\msun} = 1.3271244 $\times$ 10$^{20}$ m$^3$ s$^{-2}$, \rsun = 6.9566 $\times$ 10$^{8}$ m, \mj = 0.000954638698 \msun, and  \rj = 0.102792236 \rsun~ \citep{Standish:1995, Torres:2010, Eastman:2013}. All determined values for the four separate global fits are consistent with each other (within 1$\sigma$). We adopt the YY circular fit for all analysis and interpretation for KELT-14b and KELT-15b.

\begin{figure}[!ht]
\centering\epsfig{file=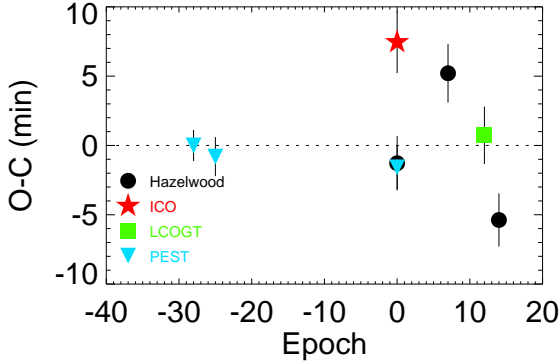,clip=,width=0.99\linewidth}
\caption{Transit time residuals for KELT-14b using our final global fit ephemeris. The times are listed in Table \ref{tbl:transitimes}.}
\label{fig:TTV}
\end{figure}

\begin{table*}
 \centering
\setlength\tabcolsep{1.5pt}
\caption{Median values and 68\% confidence interval for the physical and orbital parameters of the KELT-14 system}
  \label{tbl:KELT-14b}
  \begin{tabular}{lccccc}
  \hline
  \hline
   Parameter & Units & \textbf{Adopted Value} & Value & Value & Value \\
   & & \textbf{(YY circular)} & (YY eccentric) & (Torres circular) & (Torres eccentric)\\
 \hline
Stellar Parameters & & & & &\\
                               ~~~$M_{*}$\dotfill &Mass (\msun)\dotfill &$1.178_{-0.066}^{+0.052}$&$1.177_{-0.066}^{+0.059}$&$1.202_{-0.062}^{+0.064}$&$1.203_{-0.063}^{+0.066}$\\
                             ~~~$R_{*}$\dotfill &Radius (\rsun)\dotfill &$1.368_{-0.077}^{+0.078}$&$1.378_{-0.099}^{+0.10}$&$1.410\pm0.077$&$1.418_{-0.094}^{+0.096}$\\
                         ~~~$L_{*}$\dotfill &Luminosity (\lsun)\dotfill &$1.90_{-0.24}^{+0.28}$&$1.93_{-0.29}^{+0.34}$&$2.04_{-0.26}^{+0.29}$&$2.06_{-0.30}^{+0.34}$\\
                             ~~~$\rho_*$\dotfill &Density (cgs)\dotfill &$0.645_{-0.087}^{+0.11}$&$0.63_{-0.11}^{+0.14}$&$0.604_{-0.078}^{+0.096}$&$0.595_{-0.094}^{+0.12}$\\
                  ~~~$\log{g_*}$\dotfill &Surface gravity (cgs)\dotfill &$4.234_{-0.041}^{+0.045}$&$4.228_{-0.054}^{+0.057}$&$4.219_{-0.038}^{+0.041}$&$4.215_{-0.048}^{+0.051}$\\
                  ~~~$\teff$\dotfill &Effective temperature (K)\dotfill &$5802_{-92}^{+95}$&$5800_{-92}^{+96}$&$5815\pm88$&$5815\pm89$\\
                                 ~~~$\feh$\dotfill &Metallicity\dotfill &$0.326_{-0.089}^{+0.091}$&$0.324_{-0.089}^{+0.092}$&$0.338_{-0.085}^{+0.087}$&$0.338_{-0.085}^{+0.086}$\\
\hline
 Planet Parameters & & & & & \\
                                   ~~~$e$\dotfill &Eccentricity\dotfill &---&$0.041_{-0.027}^{+0.036}$&---&$0.039_{-0.026}^{+0.034}$\\
        ~~~$\omega_*$\dotfill &Argument of periastron (degrees)\dotfill &---&$149_{-65}^{+77}$&---&$156_{-69}^{+73}$\\
                                  ~~~$P$\dotfill &Period (days)\dotfill &$1.7100596_{-0.0000075}^{+0.0000074}$&$1.7100597\pm0.0000074$&$1.7100596_{-0.0000073}^{+0.0000074}$&$1.7100596\pm0.0000074$\\
                           ~~~$a$\dotfill &Semi-major axis (AU)\dotfill &$0.02956_{-0.00057}^{+0.00043}$&$0.02955_{-0.00057}^{+0.00048}$&$0.02976\pm0.00052$&$0.02977\pm0.00053$\\
                                 ~~~$M_{P}$\dotfill &Mass (\mj)\dotfill &$1.196\pm0.072$&$1.206_{-0.076}^{+0.079}$&$1.217_{-0.073}^{+0.075}$&$1.226_{-0.078}^{+0.081}$\\
                               ~~~$R_{P}$\dotfill &Radius (\rj)\dotfill &$1.52_{-0.11}^{+0.12}$&$1.53\pm0.14$&$1.57_{-0.11}^{+0.12}$&$1.57_{-0.13}^{+0.14}$\\
                           ~~~$\rho_{P}$\dotfill &Density (cgs)\dotfill &$0.421_{-0.083}^{+0.11}$&$0.414_{-0.091}^{+0.12}$&$0.393_{-0.076}^{+0.095}$&$0.390_{-0.082}^{+0.11}$\\
                      ~~~$\log{g_{P}}$\dotfill &Surface gravity\dotfill &$3.107_{-0.064}^{+0.066}$&$3.103_{-0.071}^{+0.075}$&$3.089\pm0.062$&$3.088_{-0.067}^{+0.068}$\\
               ~~~$T_{eq}$\dotfill &Equilibrium temperature (K)\dotfill &$1904\pm54$&$1910_{-67}^{+68}$&$1929_{-56}^{+55}$&$1934\pm65$\\
                   ~~~$\fave$\dotfill &Incident flux (\fluxcgs)\dotfill &$2.98_{-0.33}^{+0.36}$&$3.02_{-0.40}^{+0.45}$&$3.15_{-0.35}^{+0.38}$&$3.17_{-0.41}^{+0.45}$\\
 \hline
RV Parameters & & & & & \\
       ~~~$T_C$\dotfill &Time of inferior conjunction (\bjdtdb)\dotfill &$2457111.5484\pm0.0050$&$2457111.5497\pm0.0052$&$2457111.5485_{-0.0050}^{+0.0049}$&$2457111.5496\pm0.0051$\\
               ~~~$T_{P}$\dotfill &Time of periastron (\bjdtdb)\dotfill &---&$2457111.81_{-0.29}^{+0.36}$&---&$2457111.84_{-0.31}^{+0.35}$\\
                        ~~~$K$\dotfill &RV semi-amplitude (m/s)\dotfill &$179.8\pm9.0$&$181.3\pm9.3$&$179.7\pm8.9$&$181.2\pm9.2$\\
                    ~~~$M_P\sin{i}$\dotfill &Minimum mass (\mj)\dotfill &$1.177_{-0.070}^{+0.071}$&$1.185_{-0.074}^{+0.076}$&$1.196_{-0.072}^{+0.073}$\&$1.204_{-0.075}^{+0.078}$\\
                           ~~~$M_{P}/M_{*}$\dotfill &Mass ratio\dotfill &$0.000974_{-0.000051}^{+0.000052}$&$0.000981_{-0.000052}^{+0.000053}$&$0.000967_{-0.000050}^{+0.000051}$&$0.000973\pm0.000052$\\
                       ~~~$u$\dotfill &RM linear limb darkening\dotfill &$0.668\pm0.011$&$0.668\pm0.011$&$0.668\pm0.011$&$0.668\pm0.011$\\
                                 ~~~$\gamma_{AAT}$\dotfill &m/s\dotfill &$34590.0_{-6.7}^{+6.8}$&$34590.3\pm6.7$& $34590.0\pm6.7$&$34590.3\pm6.6$\\
                  ~~~$\dot{\gamma}$\dotfill &RV slope (m/s/day)\dotfill &$0.53\pm0.20$&$0.52\pm0.23$&$0.52\pm0.20$&$0.53\pm0.22$\\
                                         ~~~$\ecosw$\dotfill & \dotfill &---&$-0.019_{-0.028}^{+0.021}$&---&$-0.018_{-0.028}^{+0.020}$\\
                                         ~~~$\esinw$\dotfill & \dotfill &---&$0.005_{-0.033}^{+0.045}$&---&$0.003_{-0.032}^{+0.040}$\\
 \hline
 Primary Transit & & & & & \\
~~~$R_{P}/R_{*}$\dotfill &Radius of the planet in stellar radii\dotfill &$0.1143_{-0.0026}^{+0.0029}$&$0.1142_{-0.0026}^{+0.0029}$&$0.1141_{-0.0026}^{+0.0029}$&$0.1141_{-0.0026}^{+0.0030}$\\
           ~~~$a/R_*$\dotfill &Semi-major axis in stellar radii\dotfill &$4.64_{-0.22}^{+0.25}$&$4.60_{-0.28}^{+0.33}$&$4.54_{-0.20}^{+0.23}$&$4.51_{-0.25}^{+0.28}$\\
                          ~~~$i$\dotfill &Inclination (degrees)\dotfill &$79.67_{-0.77}^{+0.80}$&$79.5\pm1.2$&$79.36\pm0.75$&$79.2\pm1.1$\\
                               ~~~$b$\dotfill &Impact parameter\dotfill &$0.831_{-0.022}^{+0.020}$&$0.831_{-0.022}^{+0.020}$&$0.838_{-0.020}^{+0.018}$&$0.837_{-0.021}^{+0.019}$\\
                             ~~~$\delta$\dotfill &Transit depth\dotfill &$0.01306_{-0.00059}^{+0.00067}$&$0.01305_{-0.00059}^{+0.00067}$&$0.01302_{-0.00058}^{+0.00067}$&$0.01301_{-0.00058}^{+0.00068}$\\
 ~~~$T_O$\dotfill & Ephemeris from transits (\bjdtdb)\dotfill &2457091.028632$\pm$0.00047&---&---&---\\
 ~~~$P_{Transits}$\dotfill & Ephemeris period from transits (days)\dotfill & 1.7100588$\pm$0.0000025&---&---&---\\
 
                    ~~~$T_{FWHM}$\dotfill &FWHM duration (days)\dotfill &$0.0626_{-0.0025}^{+0.0018}$&$0.0626_{-0.0025}^{+0.0017}$&$0.0625_{-0.0028}^{+0.0019}$&$0.0625_{-0.0028}^{+0.0019}$\\
              ~~~$\tau$\dotfill &Ingress/egress duration (days)\dotfill &$0.0262_{-0.0037}^{+0.0046}$&$0.0261_{-0.0036}^{+0.0046}$&$0.0274_{-0.0037}^{+0.0048}$&$0.0274_{-0.0037}^{+0.0049}$\\
                     ~~~$T_{14}$\dotfill &Total duration (days)\dotfill &$0.0889_{-0.0026}^{+0.0025}$&$0.0888\pm0.0026$&$0.0900_{-0.0025}^{+0.0024}$&$0.0900\pm0.0025$\\
   ~~~$P_{T}$\dotfill &A priori non-grazing transit probability\dotfill &$0.1910_{-0.0094}^{+0.0089}$&$0.194_{-0.017}^{+0.020}$&$0.1952_{-0.0089}^{+0.0086}$ &$0.197_{-0.016}^{+0.018}$\\
             ~~~$P_{T,G}$\dotfill &A priori transit probability\dotfill &$0.240_{-0.013}^{+0.012}$&$0.244_{-0.022}^{+0.025}$&$0.246\pm0.012$&$0.248_{-0.021}^{+0.023}$\\
                   ~~~$u_{1B}$\dotfill &Linear Limb-darkening\dotfill &$0.685\pm0.026$&$0.685_{-0.026}^{+0.027}$&$0.684_{-0.025}^{+0.026}$&$0.684_{-0.025}^{+0.026}$\\
                  ~~~$u_{2B}$\dotfill &Quadratic Limb-darkening\dotfill &$0.134_{-0.021}^{+0.020}$&$0.133_{-0.021}^{+0.020}$&$0.135\pm0.020$&$0.135\pm0.020$\\
                     ~~~$u_{1I}$\dotfill &Linear Limb-darkening\dotfill &$0.294_{-0.014}^{+0.015}$&$0.294\pm0.015$&$0.293\pm0.014$&$0.292\pm0.014$\\
                  ~~~$u_{2I}$\dotfill &Quadratic Limb-darkening\dotfill &$0.2810_{-0.0075}^{+0.0074}$&$0.2810\pm0.0075$&$0.2824_{-0.0072}^{+0.0070}$&$0.2825_{-0.0073}^{+0.0071}$\\
                     ~~~$u_{1R}$\dotfill &Linear Limb-darkening\dotfill &$0.382_{-0.017}^{+0.018}$&$0.382\pm0.018$&$0.380_{-0.017}^{+0.018}$&$0.380_{-0.017}^{+0.018}$\\
                  ~~~$u_{2R}$\dotfill &Quadratic Limb-darkening\dotfill &$0.2777_{-0.010}^{+0.0096}$&$0.2776_{-0.010}^{+0.0098}$&$0.2789_{-0.0098}^{+0.0094}$&$0.2790_{-0.0099}^{+0.0093}$\\
                ~~~$u_{1Sloang}$\dotfill &Linear Limb-darkening\dotfill &$0.602_{-0.024}^{+0.025}$&$0.602_{-0.025}^{+0.026}$&$0.601_{-0.024}^{+0.025}$&$0.601_{-0.024}^{+0.025}$\\
             ~~~$u_{2Sloang}$\dotfill &Quadratic Limb-darkening\dotfill &$0.188_{-0.018}^{+0.017}$&$0.188_{-0.019}^{+0.018}$&$0.189_{-0.018}^{+0.017}$&$0.189_{-0.018}^{+0.017}$\\
                     ~~~$u_{1V}$\dotfill &Linear Limb-darkening\dotfill &$0.484\pm0.021$&$0.484_{-0.021}^{+0.022}$&$0.483_{-0.020}^{+0.021}$&$0.483_{-0.020}^{+0.021}$\\
                  ~~~$u_{2V}$\dotfill &Quadratic Limb-darkening\dotfill &$0.247_{-0.014}^{+0.013}$&$0.247_{-0.014}^{+0.013}$&$0.248\pm0.013$&$0.248_{-0.014}^{+0.013}$\\
\hline
Secondary Eclipse & & & & & \\
                  ~~~$T_{S}$\dotfill &Time of eclipse (\bjdtdb)\dotfill &$2457110.6934\pm0.0050$&$2457112.384_{-0.029}^{+0.022}$&$2457110.6935_{-0.0050}^{+0.0049}$&$2457112.384_{-0.029}^{+0.022}$\\
                           ~~~$b_{S}$\dotfill &Impact parameter\dotfill &---&$0.842_{-0.063}^{+0.080}$&---&$0.845_{-0.059}^{+0.071}$\\
                  ~~~$T_{S,FWHM}$\dotfill &FWHM duration (days)\dotfill &---&$0.0609_{-0.019}^{+0.0065}$&---&$0.0612_{-0.019}^{+0.0069}$\\
            ~~~$\tau_S$\dotfill &Ingress/egress duration (days)\dotfill &---&$0.0275_{-0.0071}^{+0.013}$&---&$0.0286_{-0.0071}^{+0.013}$\\
                   ~~~$T_{S,14}$\dotfill &Total duration (days)\dotfill &---&$0.0872_{-0.0048}^{+0.0035}$&---&$0.0887_{-0.0045}^{+0.0034}$\\
   ~~~$P_{S}$\dotfill &A priori non-grazing eclipse probability\dotfill &---&$0.1910_{-0.0094}^{+0.0091}$&---&$0.1953_{-0.0090}^{+0.0088}$\\
             ~~~$P_{S,G}$\dotfill &A priori eclipse probability\dotfill &---&$0.240\pm0.013$&---&$0.246\pm0.012$\\
 \hline
 \hline
\end{tabular}
\end{table*}

\begin{table*}
 \centering
\setlength\tabcolsep{1.5pt}
\caption{Median values and 68\% confidence interval for the physical and orbital parameters of the KELT-15 system}
  \label{tbl:KELT-15b}
  \begin{tabular}{lccccc}
  \hline
  \hline
   Parameter & Units & \textbf{Adopted Value} & Value & Value & Value \\
   & & \textbf{(YY circular)} & (YY eccentric) & (Torres circular) & (Torres eccentric)\\
 \hline
Stellar Parameters & & & & &\\
                               ~~~$M_{*}$\dotfill &Mass (\msun)\dotfill &$1.181_{-0.050}^{+0.051}$&$1.218_{-0.071}^{+0.10}$&$1.216_{-0.055}^{+0.057}$&$1.244_{-0.074}^{+0.092}$\\
                             ~~~$R_{*}$\dotfill &Radius (\rsun)\dotfill &$1.481_{-0.041}^{+0.091}$&$1.63_{-0.18}^{+0.30}$&$1.493_{-0.042}^{+0.082}$&$1.60_{-0.17}^{+0.34}$\\
                         ~~~$L_{*}$\dotfill &Luminosity (\lsun)\dotfill &$2.58_{-0.20}^{+0.35}$&$3.11_{-0.69}^{+1.3}$&$2.65_{-0.20}^{+0.32}$&$3.04_{-0.65}^{+1.4}$\\
                             ~~~$\rho_*$\dotfill &Density (cgs)\dotfill &$0.514_{-0.076}^{+0.034}$&$0.40_{-0.14}^{+0.15}$&$0.518_{-0.071}^{+0.032}$&$0.42_{-0.17}^{+0.15}$\\
                  ~~~$\log{g_*}$\dotfill &Surface gravity (cgs)\dotfill &$4.168_{-0.044}^{+0.019}$&$4.100_{-0.11}^{+0.086}$&$4.174_{-0.040}^{+0.018}$&$4.120_{-0.14}^{+0.084}$\\
                  ~~~$\teff$\dotfill &Effective temperature (K)\dotfill &$6003_{-52}^{+56}$&$6017_{-57}^{+58}$&$6021_{-61}^{+60}$&$6021_{-60}^{+61}$\\
                                 ~~~$\feh$\dotfill &Metallicity\dotfill &$0.047\pm0.032$&$0.051_{-0.032}^{+0.033}$&$0.051_{-0.033}^{+0.034}$&$0.051\pm0.033$\\
 & & & & & \\
 \hline
 Planet Parameters & & & & & \\
                                   ~~~$e$\dotfill &Eccentricity\dotfill &---&$0.132_{-0.090}^{+0.13}$&---&$0.133_{-0.091}^{+0.14}$\\
        ~~~$\omega_*$\dotfill &Argument of periastron (degrees)\dotfill &---&$141_{-42}^{+71}$&---&$142_{-42}^{+76}$\\
                                  ~~~$P$\dotfill &Period (days)\dotfill &$3.329441\pm0.000016$&$3.329442\pm0.000016$&$3.329441\pm0.000016$&$3.329442\pm0.000016$\\
                           ~~~$a$\dotfill &Semi-major axis (AU)\dotfill &$0.04613\pm0.00065$&$0.04660_{-0.00092}^{+0.0013}$&$0.04657\pm0.00072$&$0.04693_{-0.00095}^{+0.0011}$\\
                                 ~~~$M_{P}$\dotfill &Mass (\mj)\dotfill &$0.91_{-0.22}^{+0.21}$&$0.94_{-0.25}^{+0.26}$&$0.93\pm0.22$&$0.95_{-0.25}^{+0.26}$\\
                               ~~~$R_{P}$\dotfill &Radius (\rj)\dotfill &$1.443_{-0.057}^{+0.11}$&$1.59_{-0.19}^{+0.31}$&$1.453_{-0.057}^{+0.098}$&$1.56_{-0.18}^{+0.34}$\\
                           ~~~$\rho_{P}$\dotfill &Density (cgs)\dotfill &$0.36_{-0.10}^{+0.11}$&$0.28_{-0.11}^{+0.15}$&$0.363_{-0.100}^{+0.11}$&$0.29_{-0.13}^{+0.16}$\\
                      ~~~$\log{g_{P}}$\dotfill &Surface gravity\dotfill &$3.02_{-0.13}^{+0.10}$&$2.95_{-0.17}^{+0.14}$&$3.03_{-0.12}^{+0.10}$&$2.96_{-0.18}^{+0.15}$\\
               ~~~$T_{eq}$\dotfill &Equilibrium temperature (K)\dotfill &$1642_{-25}^{+45}$&$1713_{-92}^{+140}$&$1645_{-25}^{+41}$&$1699_{-87}^{+150}$\\
                   ~~~$\fave$\dotfill &Incident flux (\fluxcgs)\dotfill &$1.652_{-0.100}^{+0.19}$&$1.92_{-0.37}^{+0.58}$&$1.66_{-0.10}^{+0.17}$&$1.86_{-0.35}^{+0.63}$\\
 & & & & & \\
 \hline
RV Parameters & & & & & \\
       ~~~$T_C$\dotfill &Time of inferior conjunction (\bjdtdb)\dotfill &$2457029.1663_{-0.0073}^{+0.0078}$&$2457029.1691_{-0.0081}^{+0.0083}$&$2457029.1663_{-0.0073}^{+0.0079}$&$2457029.1688_{-0.0080}^{+0.0084}$\\
               ~~~$T_{P}$\dotfill &Time of periastron (\bjdtdb)\dotfill &---&$2457029.49_{-0.29}^{+0.64}$&---&$2457029.50_{-0.29}^{+0.71}$\\
                        ~~~$K$\dotfill &RV semi-amplitude (m/s)\dotfill &$110\pm26$&$113\pm30$&$110\pm26$&$113_{-29}^{+30}$\\
                    ~~~$M_P\sin{i}$\dotfill &Minimum mass (\mj)\dotfill &$0.91_{-0.22}^{+0.21}$&$0.94_{-0.25}^{+0.26}$&$0.93\pm0.22$&$0.95_{-0.25}^{+0.26}$\\
                           ~~~$M_{P}/M_{*}$\dotfill &Mass ratio\dotfill &$0.00073\pm0.00017$&$0.00073\pm0.00019$&$0.00073\pm0.00017$&$0.00073\pm0.00019$\\
                       ~~~$u$\dotfill &RM linear limb darkening\dotfill &$0.6290_{-0.0058}^{+0.0062}$&$0.6275_{-0.0059}^{+0.0065}$&$0.6276_{-0.0060}^{+0.0066}$&$0.6272_{-0.0060}^{+0.0066}$\\
                                 ~~~$\gamma_{AAT}$\dotfill &m/s\dotfill &$12204_{-19}^{+18}$&$12203\pm21$&$12204\pm19$&$12204\pm20$\\
                             ~~~$\gamma_{CORALIE}$\dotfill &m/s\dotfill &$12216\pm22$&$12212\pm22$&$12216\pm21$&$12211\pm22$\\
                                         ~~~$\ecosw$\dotfill & \dotfill &---&$-0.073_{-0.10}^{+0.073}$&---&$-0.074_{-0.10}^{+0.074}$\\
                                         ~~~$\esinw$\dotfill & \dotfill &---&$0.050_{-0.082}^{+0.14}$&---&$0.042_{-0.085}^{+0.16}$\\
 & & & & & \\
 \hline
 Primary Transit & & & & & \\
~~~$R_{P}/R_{*}$\dotfill &Radius of the planet in stellar radii\dotfill &$0.1001_{-0.0021}^{+0.0022}$&$0.1005_{-0.0023}^{+0.0025}$&$0.1001_{-0.0020}^{+0.0021}$&$0.1001_{-0.0021}^{+0.0022}$\\
           ~~~$a/R_*$\dotfill &Semi-major axis in stellar radii\dotfill &$6.70_{-0.35}^{+0.14}$&$6.16_{-0.83}^{+0.68}$&$6.72_{-0.32}^{+0.13}$&$6.29_{-0.99}^{+0.67}$\\
                          ~~~$i$\dotfill &Inclination (degrees)\dotfill &$88.3_{-1.7}^{+1.2}$&$87.8_{-2.3}^{+1.6}$&$88.4_{-1.6}^{+1.1}$&$88.1_{-2.2}^{+1.3}$\\
                               ~~~$b$\dotfill &Impact parameter\dotfill &$0.20_{-0.14}^{+0.18}$&$0.22_{-0.15}^{+0.19}$&$0.19_{-0.13}^{+0.17}$&$0.20_{-0.13}^{+0.17}$\\
                             ~~~$\delta$\dotfill &Transit depth\dotfill &$0.01003_{-0.00041}^{+0.00044}$&$0.01009_{-0.00046}^{+0.00050}$&$0.01001_{-0.00040}^{+0.00043}$&$0.01002_{-0.00041}^{+0.00044}$\\
                    ~~~$T_{FWHM}$\dotfill &FWHM duration (days)\dotfill &$0.1552_{-0.0016}^{+0.0015}$&$0.1552_{-0.0018}^{+0.0017}$&$0.1551_{-0.0016}^{+0.0015}$&$0.1550\pm0.0016$\\
              ~~~$\tau$\dotfill &Ingress/egress duration (days)\dotfill &$0.01635_{-0.00079}^{+0.0021}$&$0.01656_{-0.00093}^{+0.0026}$&$0.01627_{-0.00074}^{+0.0019}$&$0.01631_{-0.00076}^{+0.0020}$\\
                     ~~~$T_{14}$\dotfill &Total duration (days)\dotfill &$0.1719_{-0.0021}^{+0.0025}$&$0.1722_{-0.0024}^{+0.0030}$&$0.1717_{-0.0020}^{+0.0024}$&$0.1717_{-0.0021}^{+0.0025}$\\
   ~~~$P_{T}$\dotfill &A priori non-grazing transit probability\dotfill &$0.1343_{-0.0027}^{+0.0072}$&$0.155_{-0.026}^{+0.058}$&$0.1340_{-0.0026}^{+0.0066}$&$0.151_{-0.025}^{+0.068}$\\
             ~~~$P_{T,G}$\dotfill &A priori transit probability\dotfill &$0.1642_{-0.0036}^{+0.0092}$&$0.190_{-0.032}^{+0.070}$&$0.1637_{-0.0033}^{+0.0084}$&$0.185_{-0.031}^{+0.084}$\\
                     ~~~$u_{1I}$\dotfill &Linear Limb-darkening\dotfill &$0.2500_{-0.0069}^{+0.0078}$&$0.2471_{-0.0079}^{+0.0086}$&$0.2482_{-0.0073}^{+0.0083}$&$0.2468_{-0.0077}^{+0.0086}$\\
                  ~~~$u_{2I}$\dotfill &Quadratic Limb-darkening\dotfill &$0.2964_{-0.0036}^{+0.0027}$&$0.2980_{-0.0040}^{+0.0036}$&$0.2972_{-0.0037}^{+0.0028}$&$0.2980_{-0.0039}^{+0.0034}$\\
                     ~~~$u_{1R}$\dotfill &Linear Limb-darkening\dotfill &$0.3259_{-0.0080}^{+0.0092}$&$0.3231_{-0.0086}^{+0.0098}$&$0.3237_{-0.0083}^{+0.0098}$&$0.3227_{-0.0085}^{+0.0098}$\\
                  ~~~$u_{2R}$\dotfill &Quadratic Limb-darkening\dotfill &$0.3027_{-0.0046}^{+0.0033}$&$0.3042_{-0.0048}^{+0.0038}$&$0.3037_{-0.0048}^{+0.0035}$&$0.3043_{-0.0048}^{+0.0037}$\\
                     ~~~$u_{1V}$\dotfill &Linear Limb-darkening\dotfill &$0.4158_{-0.0091}^{+0.011}$&$0.4132_{-0.0092}^{+0.011}$&$0.4133_{-0.0094}^{+0.011}$&$0.4127_{-0.0094}^{+0.011}$\\
                  ~~~$u_{2V}$\dotfill &Quadratic Limb-darkening\dotfill &$0.2858_{-0.0061}^{+0.0045}$&$0.2871_{-0.0061}^{+0.0044}$&$0.2871_{-0.0062}^{+0.0046}$&$0.2874_{-0.0061}^{+0.0045}$\\
 & & & & & \\
 \hline
Secondary Eclipse & & & & & \\
                  ~~~$T_{S}$\dotfill &Time of eclipse (\bjdtdb)\dotfill &$2457027.5016_{-0.0073}^{+0.0078}$&$2457030.68_{-0.22}^{+0.16}$&$2457027.5015_{-0.0073}^{+0.0079}$&$2457030.68_{-0.22}^{+0.16}$\\
                           ~~~$b_{S}$\dotfill &Impact parameter\dotfill &---&$0.25_{-0.17}^{+0.22}$&---&$0.22_{-0.15}^{+0.22}$\\
                  ~~~$T_{S,FWHM}$\dotfill &FWHM duration (days)\dotfill &---&$0.168_{-0.022}^{+0.046}$&---&$0.166_{-0.024}^{+0.052}$\\
            ~~~$\tau_S$\dotfill &Ingress/egress duration (days)\dotfill &---&$0.0193_{-0.0039}^{+0.0076}$&---&$0.0185_{-0.0036}^{+0.0085}$\\
                   ~~~$T_{S,14}$\dotfill &Total duration (days)\dotfill &---&$0.189_{-0.026}^{+0.051}$&---&$0.186_{-0.028}^{+0.060}$\\
   ~~~$P_{S}$\dotfill &A priori non-grazing eclipse probability\dotfill &---&$0.1394_{-0.0057}^{+0.011}$&---&$0.1383_{-0.0051}^{+0.011}$\\
             ~~~$P_{S,G}$\dotfill &A priori eclipse probability\dotfill &---&$0.1705_{-0.0072}^{+0.014}$&---&$0.1690_{-0.0064}^{+0.014}$\\

 \hline
 \hline
\end{tabular}
\end{table*}

\subsection{Transit Timing Variation Analysis}
We were careful to confirm all observation times are in the BJD$\textunderscore$TBD format \citep{Eastman:2010}. All time conversions to BJD$\textunderscore$TBD were performed in the AIJ reduction using the timestamps in the image headers. The observatory clocks from our follow-up observers are synchronised at the start of each observing session to a standard clock (atomic clock in Boulder, CO for PEST observatory) and typically the synchronization is redone through out the observing night. From our experience, we have found the time stamp in the image header and the actual start of observations can differ by a few seconds. Using only the transit timing data shown in Table \ref{tbl:transitimes} and Figure \ref{fig:TTV}, we determined a separate ephemeris from our global fit for KELT-14b. To determine an independent ephemeris, we performed a linear fit to the transit center times inferred from the global fit for each follow-up observation. With a $\chi^2$ of 26.9 and 6 degrees of freedom, we get $T_{0}$=2457091.028632$\pm$0.00046884453 (BJD$_{TDB}$) and a period of 1.7100588$\pm$0.00000247 days. The high $\chi^2$ is likely caused by systematics in the follow-up photometric observations. Although most epochs are consistent with the linear ephemeris listed (see Figure \ref{fig:TTV}), we do have a few apparent outliers ($<$10 minutes). Significant differences in measured transit times can be a result of the differences in the observatory clocks, observing procedures and conditions, and astrophysical red noise \citep{Carter:2009}. We do not see these outliers as significant. The high $\chi^2$ is likely dominated by the three transits at epoch -1, and specifically the ICO transit which differs from the PEST and Hazelwood transits by 8 minutes. However, we find no evidence of an issue with the observations time stamps and attribute the discrepancy to be a systematic and not astrophysical in nature. Therefore, we are unwilling to claim convincing evidence for significant transit timing variations for KELT-14b. With only three transits of KELT-15b, we do not attempt a TTV analysis.

\begin{table}
 \centering
 \caption{Transit times for KELT-14b.}
 \label{tbl:transitimes}
 \begin{tabular}{llllll}
    \hline
    \hline
    Epoch & $T_{C}$ 	& $\sigma_{T_{C}}$ 	& O-C &  O-C 			& Telescope \\
	  & (\bjdtdb) 	& (s) 			& (s) & ($\sigma_{T_{C}}$) 	& \\
    \hline
 -29 & 2457043.146899   & 67   &  -5.00  & -0.07 & PEST    \\
 -26 & 2457048.276707   & 83  &  -37.99  & -0.45 & PEST    \\
  -1 & 2457091.027548   & 93  & -102.01  & -1.09 & PEST    \\
  -1 & 2457091.033997  & 134  &  455.19  &  3.37 & ICO  \\
  -1 & 2457091.027674  & 121  &  -91.12 &  -0.75 & Hazelwood  \\
   6 & 2457103.002776  & 119  &  311.42  &  2.60 & Hazelwood  \\
  11 & 2457111.550157  & 169  &   57.80  &  0.34 & LCOGT  \\
  13 & 2457114.965950  & 113 &  -316.62  & -2.78 & Hazelwood  \\
   \hline
    \hline
 \end{tabular}
\end{table}

\section{False Positive Analysis}
A signal similar to a true planetary event can be created by a variety of astrophysical and non-astrophysical scenarios. As mentioned in \$\ref{highresRV}, we find no correlation between the bisector spans and the measured radial velocities (see Figure \ref{fig:BIS}). All transit depths across optical band passes are consistent and the global fit $\log{g_*}$ is consistent with the spectroscopic analysis for KELT-14. There is some discrepancy in the KELT-15 $\log{g_*}$ from the global fit and SME analysis but this is because the AAT spectra do not include the gravity$-$sensitive MgB triplet to provide a better constraint on $\log{g_*}$. All spectroscopic observations of both KELT-14 and KELT-15 were thoroughly investigated to ensure that the observed signal arises from the target star. There are no signs of multiple sets of absorption lines, and no evidence of a blended object with a similar flux as compared to the target star. Combining the agreement of $\log{g_*}$ from the global fit and SME analysis with the analysis of the stellar spectra, we can rule out all but nearby faint blended companions. Therefore, we are confident that our measured radii (see Table \ref{tbl:KELT-14b} and \ref{tbl:KELT-15b}) are not significantly underestimated. Overall, we find no evidence that KELT-14b and KELT-15b are anything other than transiting exoplanets, but a better estimate of the  $\log{g_*}$ of KELT-15 using a high-resolution spectrum covering the gravity sensitive MgB triplet would help support the planetary nature of the companion. 

We also explored the possibility of searching for line-of-sight companions by comparing photographic plates from the Palomar Observatory Sky Survey over the past $\sim$50 years \citep{Reid:1991}.  Unfortunately, the proper motions of KELT-14b and KELT-15b over the last 50 years are too small ($<$1 \arcsec) to allow us to search for the presence of background stars or place limits on line-of-sight companions.  Future adaptive optics imaging of both systems would detect the presence of any nearby faint companions, allowing us to measure any flux contamination and better constrain the planetary parameters.

\section{Discussion}
\subsection{Evolution}

As can be seen from the results of the global fit (Table \ref{tbl:KELT-14b} and \ref{tbl:KELT-15b}), KELT-14b and KELT-15b are highly inflated planets, joining the ranks of other hot Jupiters that manifest radii much larger than predicted by standard, non-irradiated models. Several authors \citep[e.g.,][]{Demory:2011} have suggested an empirical insolation threshold ($\approx 2 \times 10^8$ erg s$^{-1}$ cm$^{-2}$) above which hot Jupiters exhibit increasing amounts of radius inflation. KELT-14b and KELT-15b clearly lie above this threshold, with a current estimated insolation of 
$2.98_{-0.33}^{+0.36} \times 10^9$ erg s$^{-1}$ cm$^{-2}$ 
and
$1.652_{-0.100}^{+0.19} \times 10^9$ erg s$^{-1}$ cm$^{-2}$, 
respectively,
from the global fits, and therefore their currently large inflated radii are not surprising. At the same time, the KELT-14 and KELT-15 host stars are both found to be at present in a state of evolution wherein the stellar radii are expanding as the stars prepare to cross the Hertzsprung gap toward the red giant branch. This means that the stars' surfaces are encroaching on their planets, which presumably is in turn driving up the planets' insolations and also the rate of any tidal interactions between the planets and the stars. 

Therefore it is interesting to consider two questions. First, has KELT-14b's and KELT-15b's incident radiation from their host stars been below the empirical radius inflation threshold in the past? If either planet's insolation only recently exceeded the inflation threshold, the system could then serve as an empirical test bed for the different timescales predicted by different inflation mechanisms \citep[see, e.g.,][]{Assef:2009,Spiegel:2012}. Second, what is the expected fate of the KELT-14b and KELT-15b planets given the increasingly strong tidal interactions they are experiencing with their encroaching host stars? 

\begin{figure}
\includegraphics[angle=90,width=1\linewidth]{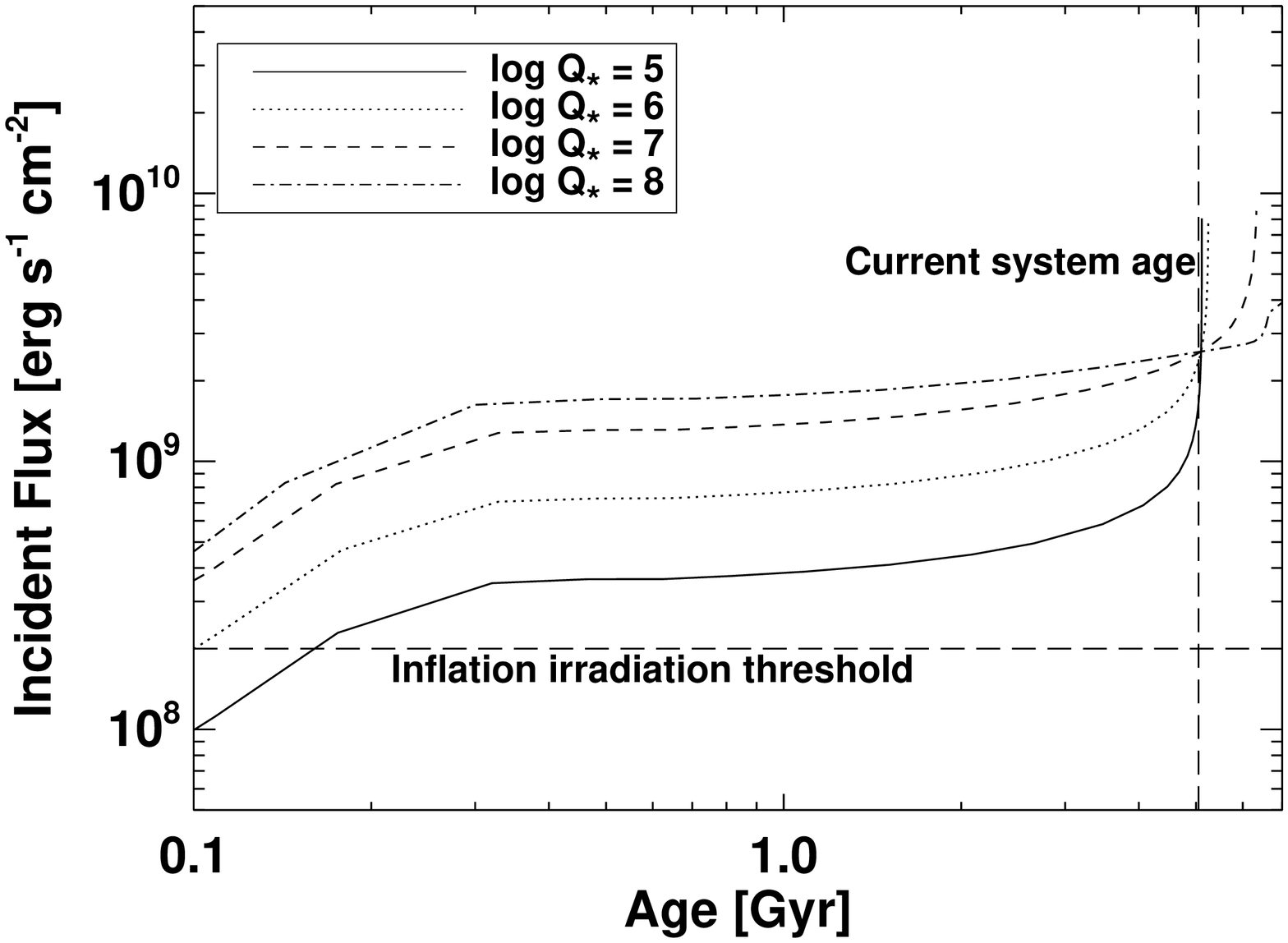}
   \vspace{-.2in}

\includegraphics[angle=90,width=1\linewidth]{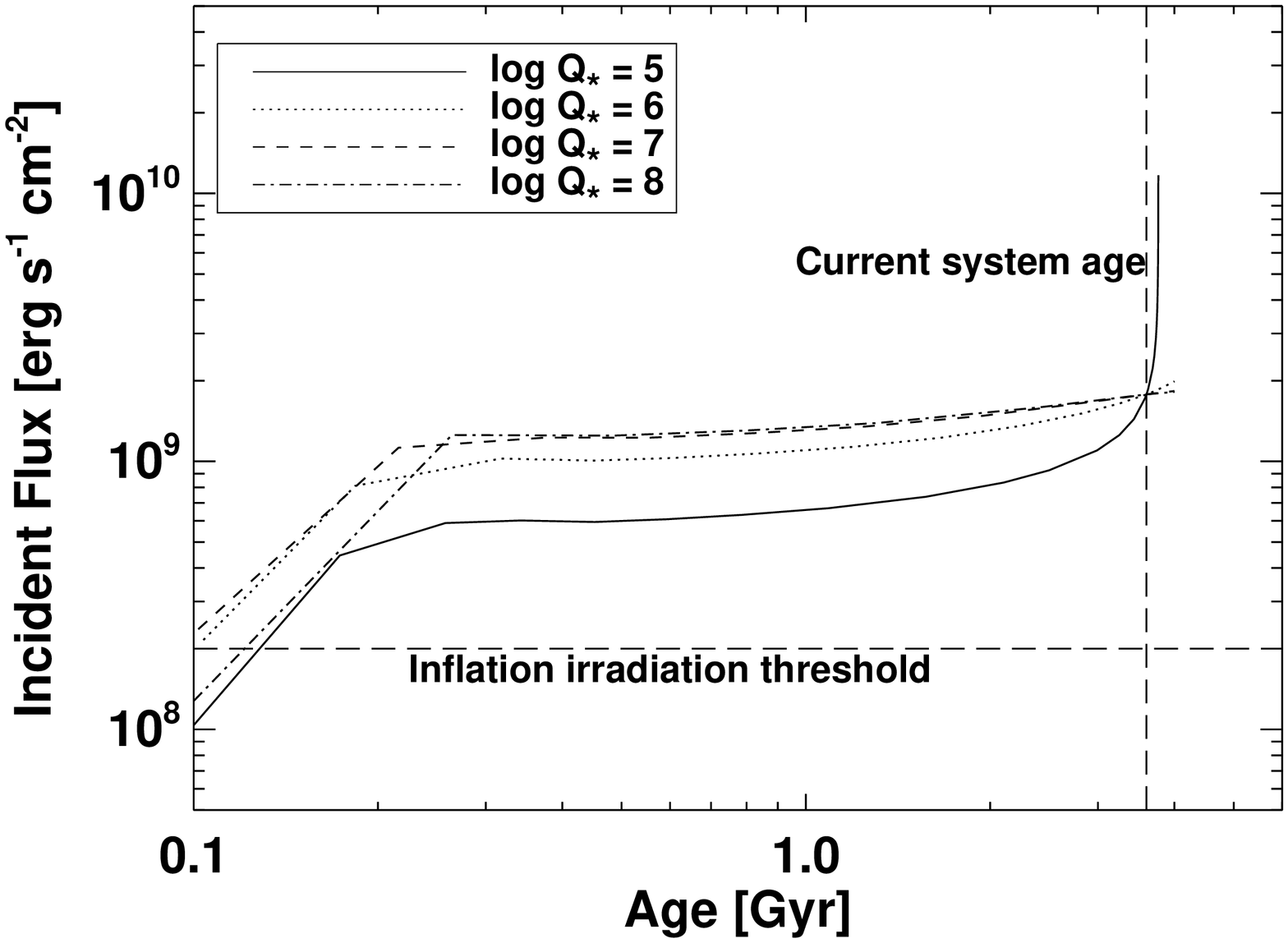}
\caption{The inflation irradiation history for (top) KELT-14 and (bottom) KELT-15 shown for test values of $\log Q'_\star$ of 5 to 8. The model assumes the stellar rotation is negligible and treats the star as a solid body. Also the model assumes a circular orbit aligned with the stellar equator. For both KELT-14b and KELT-15b, we find an the insolation received is above the empirical threshold (horizontal dashed line) determined by \citet{Demory:2011}. The vertical line represents the estimated current age of the system.}
\label{fig:Evo} 
\end{figure}

To investigate these questions, we follow \citet{Penev:2014} to simulate the reverse and forward evolution of the star-planet system, using the measured parameters listed in Table \ref{tbl:KELT-14b} and \ref{tbl:KELT-15b} as the present-day boundary conditions. This analysis is not intended to examine any type of planet-planet or planet-disk migration effects. Rather, it is a way to investigate (1) the change in insolation of the planet over time due to the changing luminosity of the star and changing star-planet separation, and (2) the change in the planet's orbital semi-major axis due to the changing tidal torque as the star-planet separation changes with the evolving stellar radius. We include the evolution of the star, assumed to follow the Yonsei-Yale stellar model with mass and metallicity. For simplicity we assume that the stellar rotation is negligible and treat the star as a solid body. We also assume a circular orbit aligned with the stellar equator throughout the analysis. The results of our simulations are shown in Figure \ref{fig:Evo}. We tested a range of values for the tidal quality factor of the star divided by the love number, $Q'_\star \equiv Q_\star / k_2$, from $\log Q'_\star = 5$ to $\log Q'_\star = 7$ (assuming a constant phase lag between the tidal bulge and the star-planet direction). 

We find that although for certain values of $Q'_\star$ the planets may have been initially below the insolation inflation threshold during the first $\sim$100 Myr, in all cases the planets have always received more than enough flux from their hosts to keep the planets irradiated beyond the insolation threshold identified by \citet{Demory:2011}. 

KELT-15b appears destined to survive for at least the next few Gyr, unless the stellar $Q'_\star$ is very small, in which case it is predicted to experience a rapid in-spiral into its host star.
In the case of KELT-14b, the current evolution of the star suggests a concomitant in-spiral of the planet over the next $\sim$1 Gyr, and even faster if the stellar $Q'_\star$ is small. This planet therefore does not appear destined to survive beyond the star's subgiant phase. As additional systems like KELT-14b are discovered and their evolution investigated in detail, it will be interesting to examine the statistics of planet survival and to compare these to predictions such as those shown here in Figure \ref{fig:Evo} to constrain mechanisms of planet-star interaction generally and the values of $Q'_\star$ specifically.

\subsection{Opportunities for Atmospheric Characterization}

\begin{figure}[!ht]
\centering\epsfig{file=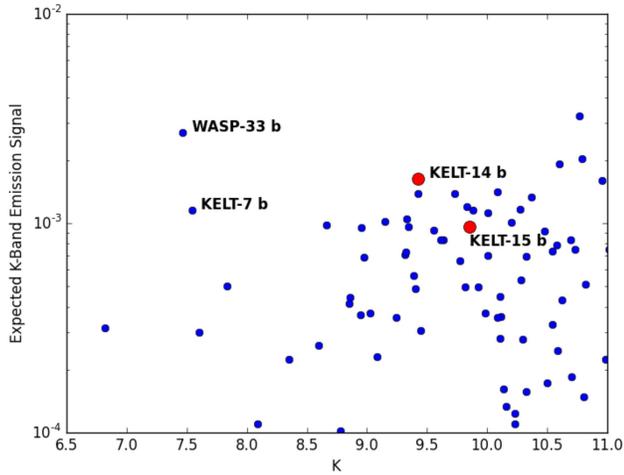,clip=,width=0.99\linewidth}
\caption{The expected day-side thermal emission from the planet in the $K$-band (assuming no redistribution of heat) for all known transiting planets brighter than a $K$-band magnitude of 11.  Along with KELT-14b and KELT-15b, we highlight WASP-33b, one of the hottest known transiting planets and KELT-7b, another very hot and very bright planet discovered by the northern component of the KELT survey.  Data are from this paper and the NASA Exoplanet Archive, accessed on 2015 August 27.}
\label{fig:Emission}
\end{figure}

Because of its very high equilibrium temperature (1904 K) and its bright $K$-band magnitude ($K$ = 9.424), KELT-14b is an excellent target for detailed atmospheric characterization.  
Specifically we note that it is an especially ideal target for eclipse observations.  Measurements during the secondary eclipse of a hot Jupiter provide a direct measurement of thermal emission from the planet's dayside and allow constraints on the connection between the atmospheric structure and climate and irradiation from the host star.  As illustrated in Figure \ref{fig:Emission}, KELT-14b has the second largest expected emission signal in the $K$-band for known transiting planets brighter than $K$ $<$ 10.5.  We therefore encourage follow up of this planet in eclipse in order to aid comparative studies of exoplanet atmospheres and better understand the connection between irradiation, albedo, and atmospheric circulation.

With an equilibrium temperature of 1642 K, KELT-15b is not as hot as KELT-14b.  However, it still has a comparably large expected emission signal in the $K$ band that should be detectable with ground-based telescopes.  Observing multiple planets in eclipse that span a range of temperatures and other properties is particularly useful for comparative exoplanetology.  

\subsection{Spectroscopic Follow-up}
From the global fit, we find that KELT-14b has an RV slope of $0.53\pm0.20$ \ms day$^{-1}$. It is possible that this RV trend is a result of a tertiary component in the system. Another discovery from the KELT survey, KELT-6b, showed an RV slope of $-0.239\pm0.037$ \ms day$^{-1}$ \citep{Collins:2014}, which was recently confirmed to be the result of KELT-6c, a 3.5 year period companion with a minimum mass of $M_P\sin{i}=3.71\pm0.21$ \mj    \space\citep{Damasso:2015}. We therefore recommend long term spectroscopic follow-up of KELT-14 to characterize the long term trend we observed. Also, the Rossiter-McLaughlin (R-M) signal for both KELT-14b and KELT-15b should be detectable with current ground$-$based facilities (The expected R-M semi-amplitude for both is $\sim$90 \ms). Two of the radial velocity observations of KELT-15b were taken during the transit and hint at a prograde orbit. However, due to the limited data acquired in transit, we do not claim KELT-15b to be in a prograde orbit but suggest future R$-$M observations to determine the spin-orbit alignment of the system.

\section{\bf Summary and Conclusions}
We present the discovery of two more transiting inflated hot Jupiter exoplanets from the KELT-South survey, KELT-14b and KELT-15b. KELT-14b, the independent discovery of WASP-122b \citep{Turner:2015} has a period of $1.7100596_{-0.0000075}^{+0.0000074}$ days, a radius of $1.52_{-0.11}^{+0.12}$~\RJ\space and a mass of $1.196\pm0.072$~\MJ. KELT-15b has a period of $3.329441\pm0.000016$ days, a radius of $1.443_{-0.057}^{+0.11}$ \RJ\space and a mass of $0.91_{-0.22}^{+0.21}$~\MJ. Additional follow-up transits are highly desirable for KELT-15b in order to better refine the ephemeris for future follow-up studies. Both KELT-14b and KELT-15b orbit host stars that are bright in the near-IR (K = 9.424 and 9.854, respectively), making them attractive targets for atmospheric characterization through secondary eclipse observations. Both should have large enough emission signals that they can be observed using ground-based observatories. These newly discovered planets increase the number of targets suitable for atmospheric characterization in the southern hemisphere.

\section*{Acknowledgements}

KELT-South is hosted by the South African Astronomical Observatory and we are grateful for their ongoing support and assistance. K.P. acknowledges support from NASA grant NNX13AQ62G. Work by B.S.G. and D.J.S. was partially supported by NSF CAREER Grant AST-1056524. Work by K.G.S. was supported by NSF PAARE grant AST-1358862. D.W. and C.G.T.'s role in this research has been supported by ARC LIEF grant LE0989347, ARC Super Science Fellowship FS100100046, and ARC Discovery grant DP130102695.

This publication makes use of data products from the Wide-field Infrared Survey Explorer, which is a joint project of the University of California, Los Angeles, and the Jet Propulsion Laboratory/California Institute of Technology, funded by the National Aeronautics and Space Administration. This publication makes use of data products from the Two Micron All Sky Survey, which is a joint project of the University of Massachusetts and the Infrared Processing and Analysis Center/California Institute of Technology, funded by the National Aeronautics and Space Administration and the National Science Foundation.

This research was made possible through the use of the AAVSO Photometric All-Sky Survey (APASS), funded by the Robert Martin Ayers Sciences Fund. This paper uses observations obtained with facilities of the Las Cumbres Observatory Global Telescope.







\bibliographystyle{apj}

\bibliography{KELT14_15}

\end{document}